\documentclass[iop]{emulateapj}

\pdfoutput=1

\usepackage{natbib}
\usepackage{color}
\usepackage{bm}
\usepackage{hyperref}
\usepackage{float}
\usepackage{latexsym}
\usepackage{amsmath}
\usepackage{amssymb}
\usepackage{multirow}
\usepackage{rotate}
\usepackage{graphicx}
\usepackage{url}
\usepackage{color}

\slugcomment{Accepted for publication in ApJ}

\begin{document}

\title{The Next Generation Virgo Cluster Survey. XXVIII. \\Characterization of the Galactic White Dwarf Population}

\shorttitle{Characterization of the Galactic White Dwarf Population in the NGVS}
\shortauthors{Fantin et al.}

\smallskip

.
\author{Nicholas J. Fantin$^{1,2}$, Patrick C{\^o}t{\'e}$^{3}$, David A. Hanes$^{2}$, S.D.J. Gwyn$^{3}$,  Luciana Bianchi$^{4}$, 
Laura Ferrarese$^{3}$, Jean-Charles Cuillandre$^{5}$, Alan McConnachie$^{3}$, Else Starkenburg$^{6}$}

\affil{$^1$Department of Physics and Astronomy, University of Victoria, Victoria, BC, V8P 1A1, Canada}
\affil{$^2$Queen's University, Department of Physics, Engineering Physics and Astronomy, Kingston, Ontario, Canada}
\affil{$^3$National Research Council of Canada, Herzberg Astronomy \& Astrophysics Program, 5071 W. Saanich Rd, Victoria, BC, V9E 2E7, Canada}
\affil{$^4$Department of Physics and Astronomy, The Johns Hopkins University, 3400 N. Charles Street, Baltimore, MD 21218, USA}
\affil{$^5$CEA/IRFU/SAp, Laboratoire AIM Paris-Saclay, CNRS/INSU, Universit\'e Paris Diderot, Observatoire de Paris, PSL Research University, F-91191 Gif-sur-Yvette Cedex, France}
\affil{$^6$Leibniz Institute for Astrophysics Potsdam (AIP), An der Sternwarte 16, D-14482 Potsdam, Germany}

\email{nfantin@uvic.ca}

\begin{abstract}

We use three different techniques to identify hundreds of white dwarf (WD) candidates in the Next Generation Virgo Cluster Survey (NGVS) based on photometry from the NGVS and GUViCS, and proper motions derived from the NGVS and the Sloan Digital Sky Survey (SDSS). Photometric distances for these candidates are calculated using theoretical color-absolute magnitude relations while effective temperatures are measured by fitting their spectral energy distributions. Disk and halo WD candidates are separated using a tangential velocity cut of 200 km~s$^{-1}$ in a reduced proper motion diagram, which leads to a sample of six halo WD candidates. Cooling ages, calculated for an assumed WD mass of 0.6$M_{\odot}$, range between 60 Myr and 6 Gyr, although these estimates depend sensitively on the adopted mass. Luminosity functions for the disk and halo subsamples are constructed and compared to previous results from the SDSS and SuperCOSMOS survey. We compute a number density of (2.81 $\pm$ 0.52) $\times 10^{-3}$~pc$^{-3}$ for the disk WD population--- consistent with previous measurements. We find (7.85 $\pm$ 4.55) $\times 10^{-6}$~pc$^{-3}$ for the halo, or 0.3\% of the disk. Observed stellar counts are also compared to predictions made by the TRILEGAL and Besan\c{c}on stellar population synthesis models. The comparison suggests that the TRILEGAL model overpredicts the total number of WDs. The WD counts predicted by the Besan\c{c}on model agree with the observations, although a discrepancy arises when comparing the predicted and observed halo WD populations; the difference is likely due to the WD masses in the adopted model halo.

\bigskip

\end{abstract}


\section{Introduction}
\label{sec:intro}

\begin{table*}	
	\caption{Searches for Halo White Dwarfs}
	\label{table:halo_wd}
		\scalebox{0.825}{
		\begin{tabular}{ccccccc}
			\hline \hline
			Reference &  Survey & Area (deg$^2$) & Bands (depth) &  Selection Parameters & Spectroscopy &Number of Halo WDs    \\    
			
			\hline
			\cite{1989LNP...328...15L}&LHS Catalog& \nodata &BVI&v$_{t}>$ 250 km~s$^{-1}$&Yes&6\\
			\cite{1997ApJ...489L.157H}&SuperCOSMOS& \nodata &BVRI&High Proper Motion&Yes&1\\
			\cite{2000ApJ...532L..41I}&proper motion survey& \nodata &B$_{J}$, R&Very High Proper Motion& Yes&2\\
			\cite{2001Sci...292..698O}& SuperCOSMOS	& 4165	&R59F (19.8), B$_{J}$ 	& RPMD& Yes	& 38\\
			\cite{2005ApJ...633L.121L}	& SUPERBLINK	& N/A	& B$_{J}$, R$_{F}$, J, H, K 	& High Proper Motion	& Yes	& 1   \\
			\cite{2005ApJ...633.1126K}&Hubble Deep Field-South& 0.0005 &F300W, F450W,&SED Fitting& \nodata &2\\
			&&&F606W(28.3), F814W&&&\\
			\cite{2006ApJ...643..402K}&New Luyten Two-Tenths& \nodata &V,J,H,K&Halo velocity ellipsoid&Yes&0\\
			\cite{2006AJ....131..571H}	& SDSS DR3	& 5282	& u,g (19.5),r,i,z	&	RPMD& Yes	& 32 (v$_{t}>$ 160 km~s$^{-1}$)     \\
			&	&	&	&	&	& 18 (v$_{t}>$ 200 km~s$^{-1}$)   \\
			Pauli et al. (2006) & SPY	& Targeted	&B $<$ 16.5	&3-D velocities (U, V, W)	& Yes	&   7 \\
			\cite{2007MNRAS.382..515V}	& SDSS Stripe 82	& 250	& $u,g,r$ (21.5),$i,z$	&	&	& 10   \\
			\cite{2008AJ....136...76H}	& SDSS DR6	& 9583	&u,g,r,i,z	& Visual Inspection 	&	&    \\
			&&&&(High Proper Motion)&Yes& 1\\
			\cite{2010ApJ...715L..21K}	& SDSS/USNO	& 2800	&$u,g,r$ ($\sim$21.0),$i,z$	& ($g-i$) = 1.5-1.75 mag	&Yes (targeted) &  3  \\
			\cite{2011MNRAS.417...93R}	& SuperCOSMOS	&~30\,000	& R59F (19.8), B$_{J}$, i$_{N}$	& RPMD ($v_t >$ 200 km~s$^{-1}$)	& No	&  93  \\
			\cite{2013ApJ...766...46H}&CFHTLS&4&$u*g'$(24.0)$r'i'z'$&RPMD&No&1\\
			\cite{2013ApJ...774...88K}&Hubble UDF& 0.0032 &I(27.0)&Proper Motion&N/A& 0 \\			
			\cite{2014AJ....147..129S}&compilation&N/A&N/A&d$<$25 pc&Yes&0\\			
			\cite{2016MNRAS.463.2453D}	&SDSS-USNO	&	3100& $u,g$ (19-22)$,r,i,z$,& RPMD (H$>$21.0,&	&    \\
			&&&J,H	&v$_{t}>$ 120 km~s$^{-1}$)& Yes& 4 \\
		    \cite{2017AJ....153...10M}	&SDSS+ Deep USNO	&	2256& $u,g,r (21.3/21.5),i,z$,& RPMD &No	&  135  \\
			\hline
			
			This Work&NGVS &104&u*,g,i,z&RPMD&No& 6 \\
			\hline
		\end{tabular}
	
}
	
	\tablecomments{RPMD = Reduced Proper Motion Diagram}
	\bigskip
	
\end{table*}

White dwarfs (WDs) represent the final evolutionary stage for stars with initial masses between $\sim$0.08 and 8$M_{\odot}$. This broad range includes $\sim$97\% of all stars, including the Sun \citep[e.g.,][]{2001PASP..113..409F}. Given this wide mass range in progenitor mass, WDs are found in virtually all stellar systems, including every major component of our Galaxy. Valuable information on the formation and evolution of the Milky Way is therefore imprinted in the properties of WDs that  we observe today \citep[e.g.,][]{2003ARA&A..41..465H}. 

WDs form as remnants of stellar cores at the end of the asymptotic giant branch phase, and their subsequent evolution is governed by a radiative cooling process that is both relatively simple and well understood \citep[e.g.,][]{2010A&ARv..18..471A}. As a result of this cooling process, the oldest WDs in the Milky Way will also be the coolest, and the WD luminosity function in any environment will, in general, show a truncation at an absolute magnitude that is determined by the age of its parent stellar system.

Despite their importance for studies of Galactic structure and stellar evolution, WDs remain, as a class, elusive. The identification of {\it halo} WDs is particularly challenging given the intrinsic faintness of old WDs and the low density of halo stars in the solar neighborhood. Table \ref{table:halo_wd} lists a number of previous surveys of field WDs, with an emphasis on those studies that aimed to identify WDs belonging to the Galactic halo. A pioneering study was the investigation of \cite{1989LNP...328...15L}, who used the Luyten Half-Second catalog \citep[LHS;][]{1979lccs.book.....L} to identify six candidate halo WDs. Roughly a decade later, \cite{2001Sci...292..698O} used the SuperCOSMOS survey to identify 38 halo WD candidates based on their Galactic space velocities, and suggested that these objects could account for 2\% or more of the dark matter in the Galactic halo. Whether these hot (and thus relatively young) objects are bona fide members of the Galactic halo, or part of the high-velocity tail of the thick disk because they lacked full 3D kinematics, has been discussed extensively in the literature \citep[e.g.,][]{2002A&ARv..11...33K,2003ApJ...586..201B,2003ARA&A..41..465H,2005ApJ...625..838B}.


About a decade ago, \cite{2006AJ....131..571H} used photometry from the Sloan Digital Sky Survey \citep[SDSS;][]{2000AJ....120.1579Y} and USNO-B proper motions to identify roughly 6000 WDs brighter than $g \sim 19.7$ and construct a luminosity function with a clear truncation at $M_{\rm bol} \sim$ 15.3. A subset of 32 high-velocity WDs --- presumably associated with the halo --- were identified by these authors. \cite{2011MNRAS.417...93R} used digitized Schmidt plates, with a limiting magnitude of $R \sim 19.5$, to search for WDs in the SuperCOSMOS Sky Survey. Adopting a proper motion selection corresponding to a tangential velocity of v$_{t} \ge$ 200 km~s$^{-1}$, these authors identified 93 possible halo WDs among a sample of $\sim$10\,000 WD candidates distributed over an area of $\sim$ 30\,000 deg$^2$.

The largest single sample of halo WDs to date was presented by \cite{2017AJ....153...10M} who obtained second epoch positions for SDSS objects covering an area of 2256 deg$^2$. From their sample of 8472 WDs, 135 were identified as belonging to the halo. This sample was used recently by \cite{2017arXiv170206984K} to estimate ages of 7.4-8.2 Gyr for the thin disk, 9.5-9.9 Gyr for the thick disk, and 12.5$^{+1.4}_{-3.4}$ Gyr for the inner Galactic halo.

Clearly, the detection of halo WDs is observationally challenging, with only $\sim$ 0.5--1\% of cataloged WDs appearing to have a halo origin. But despite their rarity, nearby halo WDs can provide interesting insights into the nature of the halo. For example, \cite{2012Natur.486...90K} used four field halo WD candidates from \citep{Pauli2006} to estimate the age of the inner halo. Adopting the initial-final mass relationship (IFMR) derived from the globular cluster M4, and combining with age estimates for the WD progenitors, \cite{2012Natur.486...90K} found an age of 11.4 $\pm$ 0.7 Gyr.

It is worth noting that, perhaps contrary to expectations, some hot WDs are expected in even the oldest stellar populations \cite[see, e.g.,][]{2011MNRAS.411.2770B,2012Natur.486...90K}. Hot WDs associated with the halo are of particular interest as they are the youngest WDs in this ancient Galactic component. These hot objects are difficult to detect with optical data alone because they can be quite faint at visible and infrared wavelengths. However, their flux is comparable to the upper main sequence in the ultraviolet (UV) bands \citep[see, e.g.,][]{2015ApJ...804...53H}, making them much easier to detect.  Furthermore, by combining UV data from the Galaxy Evolution Explorer (GALEX; \citealt{2005ApJ...619L...1M}) with optical data from the SDSS, \cite{2011MNRAS.411.2770B} showed that hot WDs can be cleanly separated from hot main-sequence stars and QSOs with high UV fluxes. On the other hand, they also noted that many of the hottest WD candidates detected by GALEX fall below the SDSS detection limits. \cite{2011MNRAS.411.2770B} used these hot, young, WDs to explore the IFMR of WDs, a key ingredient in Galactic structure models that incorporate stellar synthesis codes \citep[e.g.,][]{2003A&A...409..523R,2005A&A...436..895G}.

The Next Generation Virgo Cluster Survey (NGVS) \citep{2012ApJS..200....4F} provides homogeneous optical imaging ($u^*giz$) over a $\sim$ 100 deg$^2$ region that is roughly 3.5 magnitudes deeper than SDSS (i.e., the NGVS has a 5$\sigma$ limiting magnitude for point sources of 26.7 in the $g$-band). It thus offers high-quality optical data that can be used to characterize the WD population in this high-latitude field, particularly since deep, complementary UV imaging from GALEX also exists for the Virgo cluster region (see \S2.2). Because the NGVS is photometrically and astrometrically calibrated to the SDSS, it also provides second epoch positions for objects in the SDSS that can be used to select WD candidates from proper motions.   We note in passing that the NGVS sightline passes through a particularly interesting region of the halo --- close to the bifurcation point of the Sagittarius Stream and through the northern edge of the Virgo Overdensity. \citet{2016ApJ...819..124L} have previously used the NGVS point-source catalog to explore the properties of these halo substructures (see, e.g., their Figure 1). However, at distances of $\sim$10--50~kpc, these substructures are located well behind the comparatively local stellar populations considered here. Our WD samples  should thus be entirely representative of the disk and halo.

This paper is structured as follows. In \S\ref{sect:data}, we introduce the optical, UV, and proper motion catalogs used in our analysis, while \S\ref{sec:selection} describes three different --- and complementary --- methods for selecting WD candidates. \S\ref{sec:Properties} discusses the photometric properties of the WD candidates while \S\ref{sec:Halo} separates the disk and halo WD populations for each of our three samples. Our findings are presented in \S\ref{sec:Discussion} and conclusions are given in \S\ref{sec:conclusions}.

\smallskip
\section{The Data}
\label{sect:data}
\bigskip

In this section, we describe the optical and UV datasets used to select WD candidates in the NGVS field, including point-source identification, catalog matching, and proper motion measurements.   

\smallskip

\subsection{The Next Generation Virgo Cluster Survey (NGVS)}
\bigskip

The optical data used in this paper were obtained as part of the NGVS --- a deep, multiband ($u^*giz$) imaging survey of the Virgo cluster carried out with the Megacam instrument on the Canada France Hawaii Telescope (CFHT) between 2008 and 2013. The NGVS covers 104 deg$^2$ and extends out to the virial radii of both the Virgo A and B substructures centered, respectively, on M87 to the north and M49 to the south. Complete information on the NGVS technical details, data reduction techniques, and science goals of the NGVS may be found in \cite{2012ApJS..200....4F}.

\smallskip

Although the NGVS was primarily intended to study the properties of galaxies and other stellar systems in the Virgo cluster, the combination of its high Galactic latitude ($b \sim 75^\circ$), sub-arcsecond seeing, and long exposure times ensures that large numbers of faint disk and halo stars are detected in the survey. For instance, the NGVS has a S/N = 10 point-source depth in the $u^*$, $g$, and $i$ bands of 24.8, 25.9, and 25.1 AB mag, respectively. The seeing in all bands never exceeded 1$^{\prime\prime}$ and is particularly sharp in $g$ and $i$, with median FWHMs of 0\farcs77 and 0\farcs52, respectively \citep{2014ApJ...797..102R}. All NGVS images were processed using the MegaPipe pipeline, which uses a global background sky subtraction \citep{2008PASP..120..212G}, and were calibrated photometrically and astrometrically to the SDSS.  

\smallskip

The selection of point-source objects was performed in accordance with the method of \cite{2014ApJ...794..103D}. This technique identifies point sources based on their concentration index, $\Delta i$, which is the difference between the four- and eight-pixel diameter aperture-corrected magnitudes in the $i$~band: $\Delta{i} \equiv i_4 - i_8$. This filter is preferred for concentration measurements because, as noted above, the $i$-band imaging was obtained under the best seeing conditions. For point sources, the measured concentration indices should be centered on zero with minimal scatter; increasing the aperture diameter on extended sources will lead to brighter magnitudes and thus to a larger $\Delta{i}$. Note that a negative concentration index can occur when the local background is overestimated. 

\begin{figure}[!t]
	
\includegraphics[angle=0,width=.5\textwidth]{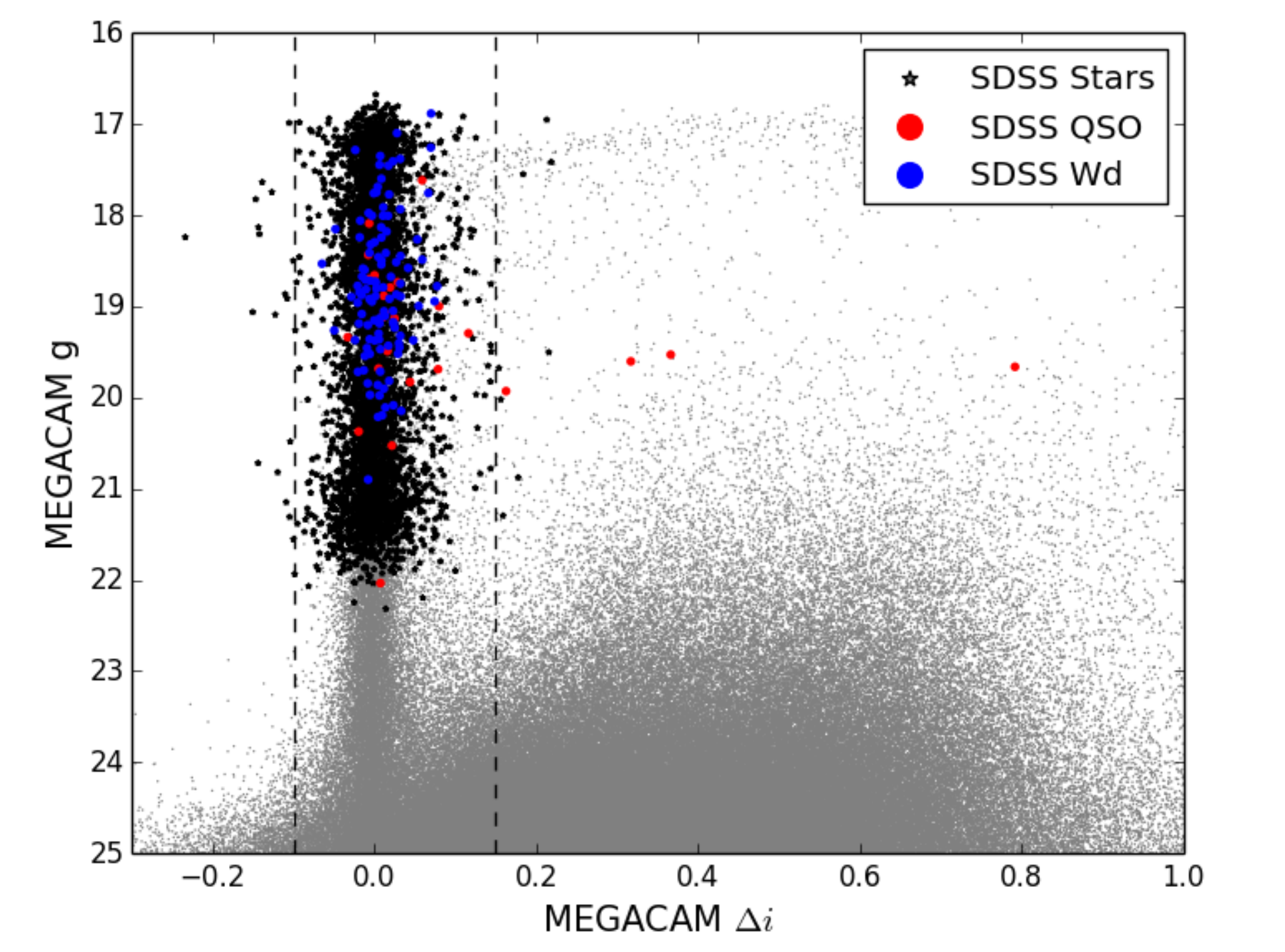}
\caption{Concentration index, $\Delta{i}$, vs. $g$-band magnitude for NGVS sources (gray) and stellar objects from SDSS (black). A narrow stellar sequence centered on $\Delta{i}$ = 0 is apparent, along with a broad cloud of background galaxies at fainter $g$-band magnitudes. Only 100,000 point sources from the NGVS are plotted for clarity. The dashed lines show our selection window for NGVS point sources, consistent with \cite{2014ApJ...794..103D}. 
	\bigskip }
\label{fig:concentration_index}
\end{figure}

In order to quantify the range of concentration index needed to extract point sources, a set of spectroscopically classified stars was queried from the SDSS. A plot of concentration index versus $g$-band magnitude is shown in Figure \ref{fig:concentration_index}, where the black stars are stellar sources from the SDSS, red circles are QSOs from \cite{2014A&A...563A..54P}, and blue circles are spectroscopically confirmed WDs from \cite{2013ApJS..204....5K}. The dashed lines show the range in concentration index used by \cite{2014ApJ...794..103D} to identify point sources: 
\begin{equation}
\begin{array}{ccccc}
    	-0.1 & \le & \Delta{i} & \le  & +0.15. \\
\end{array}
\end{equation}
Figure \ref{fig:concentration_index} shows that this selection is in excellent agreement with the location of the SDSS-selected point sources, including 99.3\% of SDSS stars. Applying this selection on concentration index to the NGVS catalog returns 5.3 million point-like objects brighter than $g$ = 24.5. 

Note that Figure~\ref{fig:concentration_index} also
shows the saturation limit of the NGVS long exposures ($g \sim$ 17.5) as noted in \cite{2015ApJ...812...34L}. The saturation causes photons to spill into adjacent pixels, which in turn leads to a more extended object, and hence a trail scattering toward positive $\Delta i$ values.

\smallskip
\subsection{The GALEX Ultraviolet Virgo Cluster Survey (GUViCS)}
\bigskip

The UV point-source catalog used in this study is based on the GALEX Ultraviolet Virgo Cluster Survey \citep[GUViCS;][]{2011A&A...528A.107B} catalog of \citet{2014A&A...569A.124V}, which includes data for 1.2 million point sources. The GUViCS survey itself, which consists of UV photometry in both the near-UV ($\lambda_{\rm eff}$ = 2316~\AA) and far-UV ($\lambda_{\rm eff}$ = 1539~\AA) channels of GALEX, is an amalgamation of imaging from multiple programs, including the All Sky Imaging Survey (AIS), the Medium Imaging Survey (MIS), the Deep Imaging Survey (DIS), the Nearby Galaxies Survey (NGS) and various GALEX PI programs (see \citealt{2007ApJS..173..682M} for details on these surveys). Although the Virgo cluster region was fully covered by the AIS early in the mission lifetime, the GUViCS team was awarded time in 2010 to cover the NGVS footprint to a depth equivalent to that of the MIS. In all, GUViCS spans an area of $\sim$~120 deg$^2$ centered on M87 and covers most of the NGVS field, although some regions were avoided due to the presence of bright stars that would have saturated the GALEX detectors. 

To maximize the point-source depth, only those detections with the highest signal-to-noise ratio for a given object were retained. The final catalog contains the updated NUV photometry to a depth of $m_{\rm NUV} \simeq $ 23.1 mag, and the previously acquired AIS FUV photometry to a depth of $m_{\rm FUV} \simeq$ 19.9 mag. Only the deeper NUV photometry was used to select candidates because just 41\% of objects were detected in the FUV. For reference, the instrumental resolution in the NUV band is $\sim$ 4\arcsec--6\arcsec, with a median of 5\farcs3 \citep[e.g.,][]{2014AdSpR..53..900B}.

\subsubsection{Bright Stellar Sources}

In assembling the GUViCS point-source catalog, \cite{2014A&A...569A.124V} removed 12\,211 bright foreground stars that appeared in the SIMBAD database. This decision was appropriate given that their immediate science drivers were extragalactic in nature. Here, however, we are interested in matching the optical and UV photometry for a highly complete sample of stars, so the GUViCS  catalog was combined with the hot star catalog of \cite{2011MNRAS.411.2770B}.

\subsubsection{Multiple Matches}

When matching the catalogs, there are two situations in which a multiple match can occur. The first, and most common, case is when multiple optical counterparts are attributed to a single GUViCS source. Indeed, the differences in spatial resolution and survey depth between the two catalogs mean that a number of NGVS objects are often matched to a single GUViCS object. This is not a rare occurrence given the depth of the NGVS data. The second, more infrequent, case is when multiple GUViCS sources are attributed to a single NGVS source. In either case, multiple matches are discarded because the optical-UV colors are compromised and must be excluded from our analysis. In all, matching the point-source catalogs from NGVS and GUViCS leads to 104\,050 unique matches. 

\subsubsection{Spurious Matches}

The high spatial density of the optical data means that there is a possibility that two unrelated sources may be matched (i.e, a spurious match). The likelihood of spurious matches was estimated in two ways. First, a collection of cutouts of one square degree from the GUViCS sample were selected, and one degree was added to both their right ascension and declination. The resulting tessellated cutout was then matched to the NGVS catalog and the spurious match rate was calculated as the total matches divided by the total number of points within the square degree offset. Repeating this exercise for five different pointings gives a spurious match rate of $\sim$2\%

\smallskip

A second method to estimate the spurious match rate used a Monte Carlo approach. A 0.5 deg$^2$ field containing both GUViCS and NGVS data was used to assess the spatial density of both catalogs. The NGVS data were first replaced by an equal number of randomly generated coordinates using the \tt{random.uniform} \normalfont function in Python. A nearest-neighbors algorithm was then applied to the resulting coordinates in order to determine the distance to the three nearest mock optical objects for each GUViCS object. If the distance to one of the three nearest neighbors was less than three arcseconds, then it was considered to be a match. This exercise was repeated 100 times for each GUViCS point. The average spurious match rate found in this way is $\sim$3\%. 

\smallskip

These methods suggest that the contamination from spurious matches is roughly 2--3\%. Of course, the actual contamination rate will be lower than this because the colors resulting from spurious matches are likely to be very odd and inconsistent between different indices. This was tested explicitly for the WDs selected from the NGVS-GUViCS catalog by matching 100\,000 GUViCS objects to random NGVS objects; in this case, only 0.27\% of matched objects would have been identified as candidate WDs. Scaling this to the number of objects in the NGVS-GUViCS matched sample shows that we expect less than one spurious match to be identified as a WD candidate.

\smallskip
\subsection{Proper Motions: The Sloan Digital Sky Survey}

\begin{figure*}[!t]
	\includegraphics[angle=0,width=.99\textwidth]{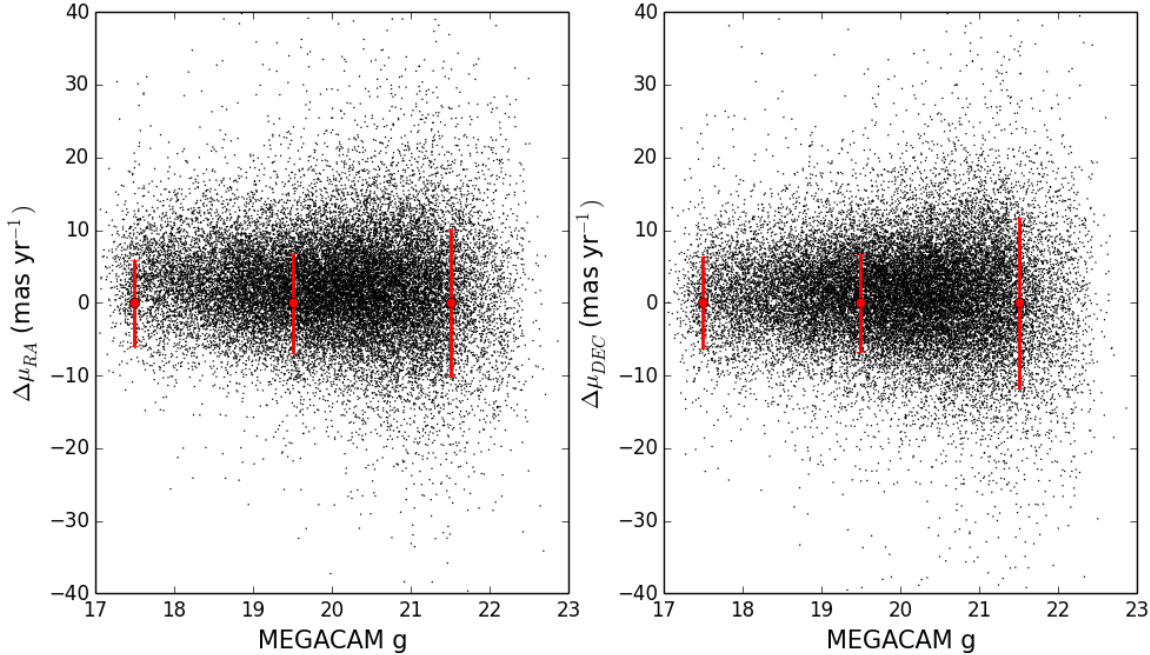}
	\caption{Comparison of proper motions derived from the NGVS and SDSS (abscissa) to those from the USNO-B catalog \citep[][ordinate]{2014AJ....148..132M}. The residuals are plotted on the y-axis as the difference between proper motion measurements. Representative error bars for three g-band magnitude bins (17, 19, and 21) are also plotted. }
	\label{fig:USNO}
	\bigskip
\end{figure*}

\bigskip

Proper motions for objects in the NGVS were calculated by comparing their coordinates with those obtained from the SDSS DR7 \citep{2009ApJS..182..543A}. This provides the longest possible baseline, since subsequent SDSS data releases used more recent observations \citep{2014AJ....148..132M}. The elapsed time between the SDSS and NGVS imaging ranges between 3 and 9 years, with an average of $\sim$ 7 years.

Deriving positions and epochs for the NGVS is not entirely straightforward because the final NGVS images were created by stacking individual frames with relatively short exposure times, whereas the SDSS images are composed of single draft scans. In some cases, the NGVS data acquisition process stretched over a period of a few years; in those instances, objects with high proper motions can be smeared in the direction of motion. In order to mitigate errors in positions caused by this effect, we restricted our analysis to only those fields in which all images were acquired within a single observing season.

\subsubsection{Comparison to the USNO Catalog}

In order to check the accuracy of the NGVS-SDSS proper motions, our measurements were compared to those from the United States Naval Observatory (USNO) catalog \citep{2014AJ....148..132M}, which partially overlaps with the NGVS. The USNO proper motions were computed from SDSS positions and follow-up observations obtained with the Steward Observatory Bok 90 inch telescope. Their images were acquired in the $r$-band with an average baseline of six years; the quoted statistical uncertainties on the proper motions range between 5 mas~yr$^{-1}$ for brighter objects ($r \sim$ 18) and 15 mas~yr$^{-1}$ for the faintest objects ($r \sim$ 22). The USNO catalog, which is complete to $r \sim$ 22 ($\delta < 8.5$),  is the deepest proper motion survey currently available in the NGVS footprint. 

\smallskip

A comparison between the proper motions derived in this work and those from the USNO catalog is shown in Figure \ref{fig:USNO}.Also plotted are representative error bars for three g-band magnitude bins in order to show the magnitude dependence of the derived proper motion errors. At bright magnitudes the resulting average error is $\sim$6 mas yr$^{-1}$ in both RA and DEC, and this increases to $\sim$12 mas yr$^{-1}$ in the faintest bin.

\subsubsection{Proper Motion Errors}

To quantify the uncertainties associated with the method described above, a sample of spectroscopically confirmed QSOs from the SDSS was compiled from \cite{2014A&A...563A..54P} over the magnitude range shown in \ref{fig:USNO}. QSOs are point sources but are, of course, extragalactic in nature and so have negligible proper motions. The mean proper motions of the QSOs within the NGVS field in right ascension and declination are 3.3$\pm$0.4 and 1.2$\pm$0.3 mas~yr$^{-1}$, respectively. However, these errors are rather small compared to the range of proper motion examined in this paper and thus should not affect our findings. In our analysis, only those objects having combined SDSS positional errors of less than 0$^{\prime\prime}$.1 were selected. With an average baseline of $\sim$ 7 years, this corresponds to a maximal uncertainty on proper motion of $\sim$ 15 mas~yr$^{-1}$. 

\section{Selection of White Dwarf Candidates}
\label{sec:selection}

In this section, we describe the methodology used to identify WD candidates in the catalogs described in \S2. Each selection method is designed to probe different regions of the WD luminosity function by applying simple color, and hence temperature, selections. Candidates were selected in three ways: (1) by using the NGVS photometry alone, which probes the hottest and youngest WDs (T$_{eff} > $ 12,500~K); (2) by combining the NGVS with UV photometry from GUViCS, which selects objects over a wider temperature range (T$_{eff} > $ 9,500~K); and (3) by using proper motions derived from the NGVS and SDSS, which allows for a selection over all temperatures, but is limited to sources with first epoch positions needed for proper motion measurements. We conclude with a brief discussion of possible contamination by other point-like objects.

\smallskip
\subsection{Method 1: Selection from the NGVS Color-Color Diagram}
\bigskip

\begin{figure}[!t]
	\includegraphics[angle=0, width=.49\textwidth]{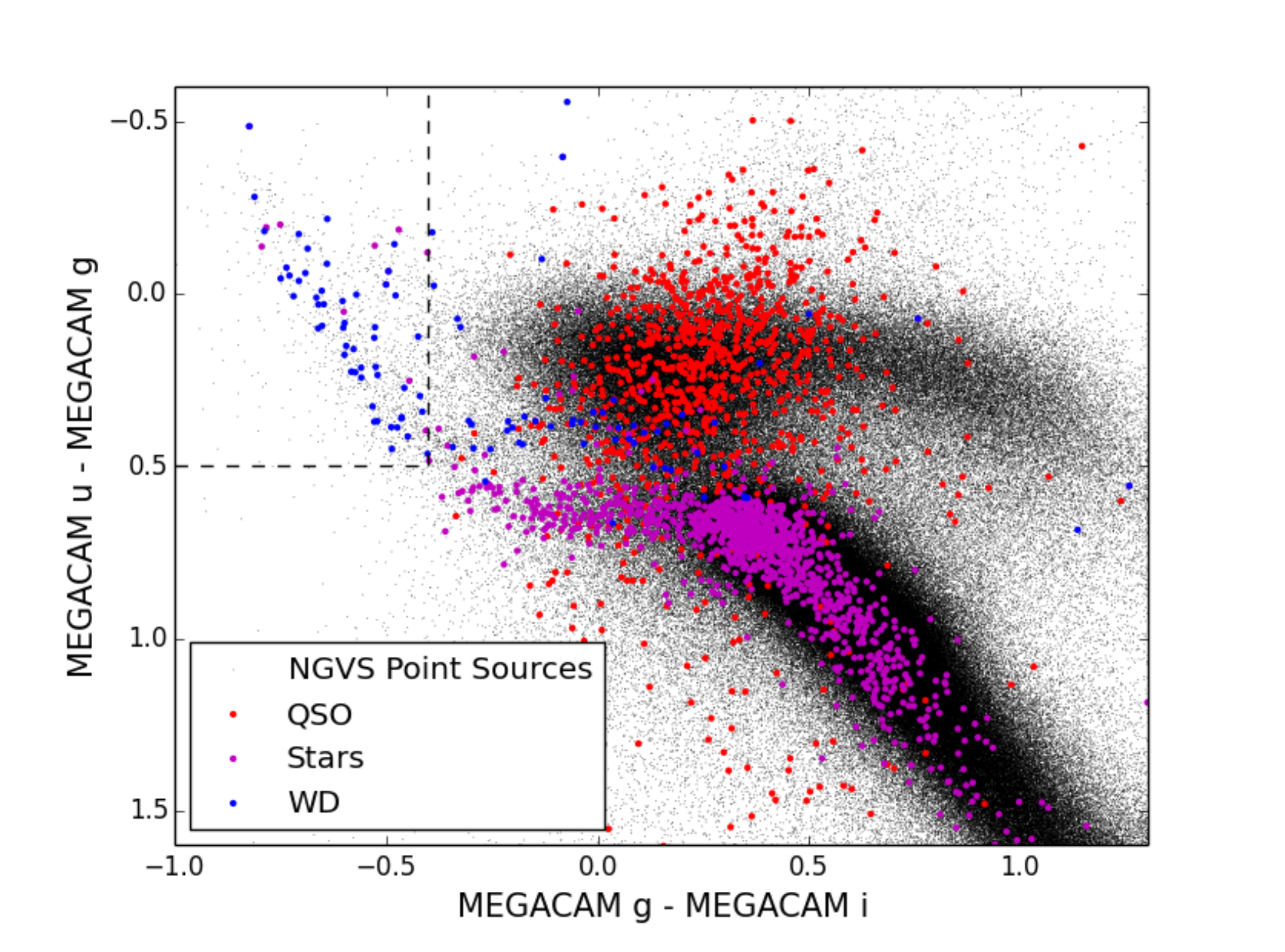}
	\caption{Color-color diagram of point sources in the NGVS (black dots). Also shown are WDs (blue), QSOs (red), and stars (magenta) selected from the SDSS inside the NGVS footprint. The dashed lines indicate our selection region for candidate WDs in the NGVS based on the sample of spectroscopically confirmed WDs from \cite{2013ApJS..204....5K}. }
	\label{fig:NGVScc}
	\bigskip
\end{figure}

The identification of WD candidates from the NGVS was performed using a color-color selection in the ($u^*-g$), ($g-i$) plane. Such a color-color diagram can be seen in Figure \ref{fig:NGVScc}, which includes spectroscopically confirmed WDs \citep[blue;][]{2013ApJS..204....5K}, QSOs \citep[red;][]{2014A&A...563A..54P}, and stars from SDSS DR7 \citep[magenta;][]{2009ApJS..182..543A}. 

\begin{figure*}[!t]
	\includegraphics[angle=0,width=.49\textwidth]{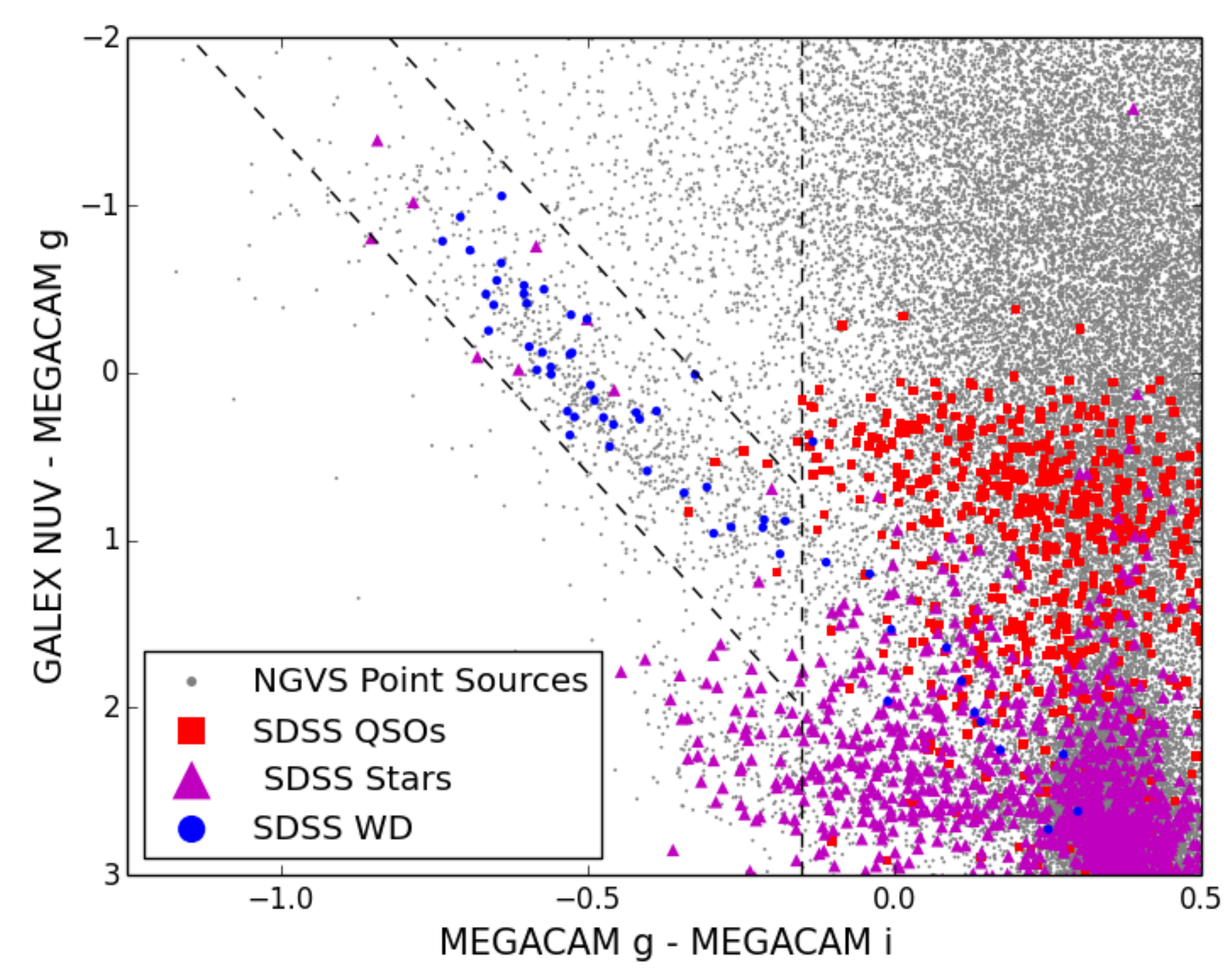}
	\includegraphics[angle=0,width=.49\textwidth]{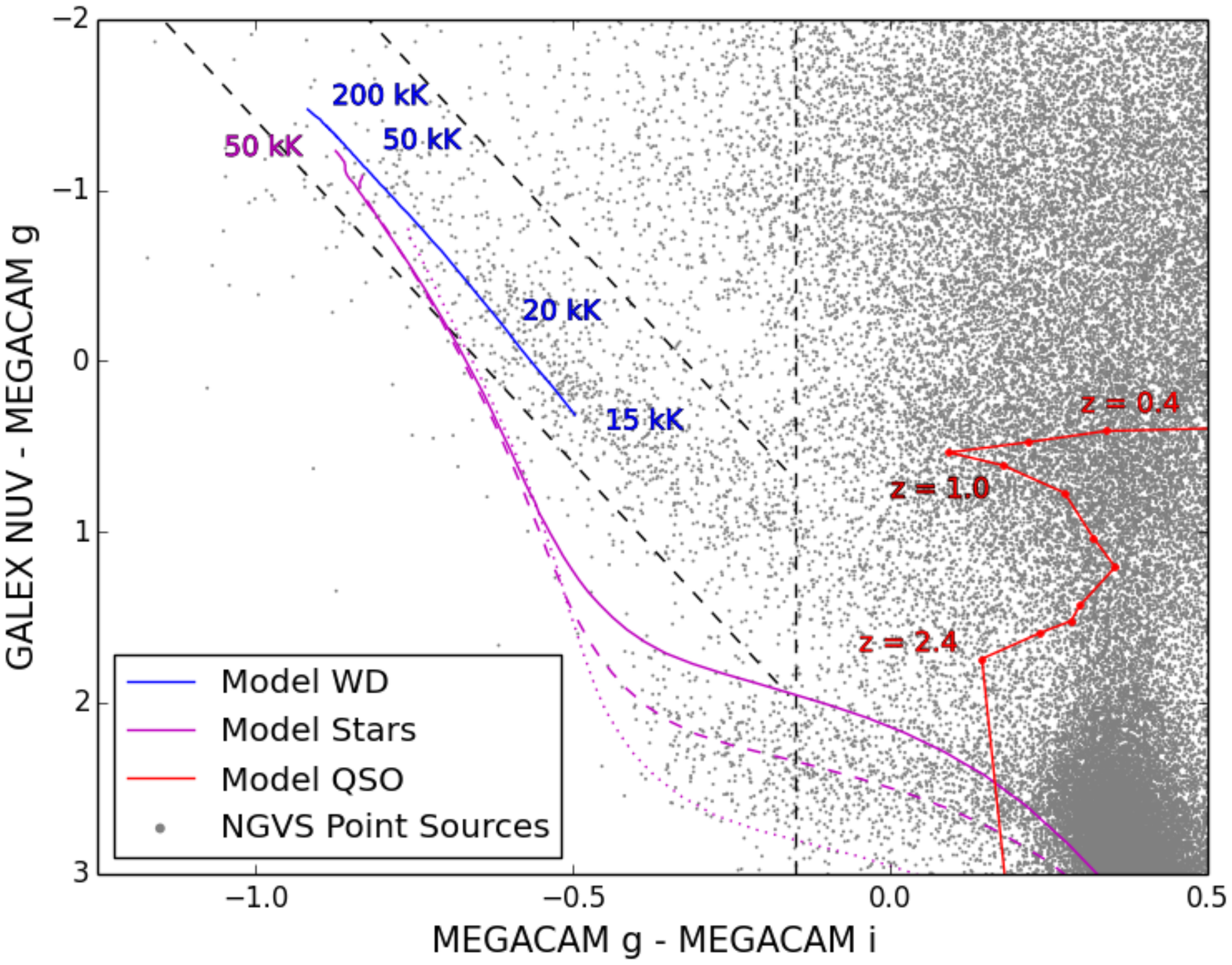}
	
	\caption{ ($g - i$, NUV -- $g$) color-color diagram for the matched NGVS-GUViCS point-source objects (gray). Left: Spectroscopically confirmed SDSS WDs (blue circles), QSOs (red squares), and stars (magenta triangles) are shown in relation to the NGVS-GUViCS objects. Right: Model tracks  show the expected locations of WDs (blue) \citep{HL95}, QSOs (red) \citep{2009AJ....137.3761B}, and for stars of various surface gravities (magenta) \citep{1993yCat.6039....0K}. The WD track, for log \textit{g} = 8.0, is plotted for temperatures between 15,000 and 200,000 K. The nonlinearity of the temperature-color relation is apparent from the identified temperatures. The model QSO colors are plotted as a function of redshift, with values between 0 and 4. Model stars with solar metallicity and log \textit{g} = 5, 4, and 3 are indicated by the solid, dashed, and dotted magenta lines, respectively. The black dashed lines indicate the color cuts applied to the NGVS-GUViCS data in order to select candidate WDs.
		\bigskip}
	\label{fig:modelcc}
\end{figure*}

Our color-color selection is shown by the black dashed lines in Figure \ref{fig:NGVScc}. This dual color selection --- ($g-i$) $\le$ 0.4 and ($u^*-g$) $\le$ 0.5 --- minimizes contamination by QSOs and main-sequence stars while maximizing the number of WDs.  The total number of WD candidates identified using this approach is 1209.

\smallskip
\subsection{Method 2: Selection from NGVS-GUViCS Color-Color Diagram}
\bigskip

WD candidates in the NGVS-GUViCS catalog were selected by applying color cuts to the point-source catalog described in \S2.2. The left hand panel of Figure~\ref{fig:modelcc} shows the location of spectroscopically confirmed WDs, QSOs, and stars from SDSS (see above). The right hand panel shows the corresponding model tracks that were specifically computed in the NGVS and GALEX colors from the grids of stellar and QSO models of \cite{2009AJ....137.3761B,2011MNRAS.411.2770B}, which include high-gravity WD models computed with the TLUSTY code \citep{HL95}, main-sequence and giant star model atmospheres computed with the Kurucz code \citep{1993yCat.6039....0K}, and various QSO templates. Comparing the left and right hand panels of Figure \ref{fig:modelcc} shows that the models and observations are generally in good agreement.

The black dashed lines in Figure \ref{fig:modelcc} show the color cuts used to select WD candidates from the NGVS and GUViCS photometry. This selection methodology was adopted after visually inspecting the locations of the spectroscopically confirmed objects in the ($g - i$), (NUV$ - g$) plane, with the goal of minimizing contamination from QSOs and hot subdwarfs (see \S\ref{sec:contaminates} for further discussion). A magnitude cut of $g < 24.5$ was also imposed to minimize contamination from background objects. Furthermore, all objects with ($g - i$) $< -0.15$ were selected, which isolates the bluest objects in the NGVS data set. An additional selection, indicated by the diagonal dashed lines, was then applied to separate the bluest main-sequence and subdwarf stars from the WDs. One final cut was imposed on the UV data by only selecting objects with uncertainties below 0.3 mag in the NUV channel. In all, these cuts result in the selection of 832 WD candidates.

A final addition to our sample of WD candidates was made using the results from \cite{2011MNRAS.411.2770B}.  A further 24 candidates were added by applying the same color cuts to the shallower GALEX AIS data, bringing the total number of candidates up to 856. Recall that this step is required because \cite{2014A&A...569A.124V} removed bright stars in assembling the GUViCS point-source catalog. Matching these objects to SDSS DR12 reveals that 77 have spectroscopic measurements. Of these, 52 have been confirmed as WDs, nine as QSOs, and 10 as other types of hot stars, and six had S/N ratios less than four (see \S3.4 for further discussion)

\smallskip
\subsection{Method 3: Selection from NGVS-SDSS Proper Motions}
\bigskip

Our final selection method relies on the reduced proper motion diagram (RPMD) --- a distance-independent metric that can be used to separate WDs from main-sequence stars and QSOs based on their photometry and proper motions. The reduced proper motion, $H$, relates the apparent magnitude, $m$, and the proper motion, $\mu$, in arcsec~yr$^{-1}$, to the tangential velocity, $v_{t}$, in km~s$^{-1}$, and absolute magnitude, $M$:
\begin{equation}
\begin{array}{lcl}
H & = & m + 5\log\mu + 5 \\
    & = & M + 5\log v_{t} - 3.379. \\
\end{array}
\end{equation}

\begin{figure}%
	\includegraphics[angle=0,width=.49\textwidth]{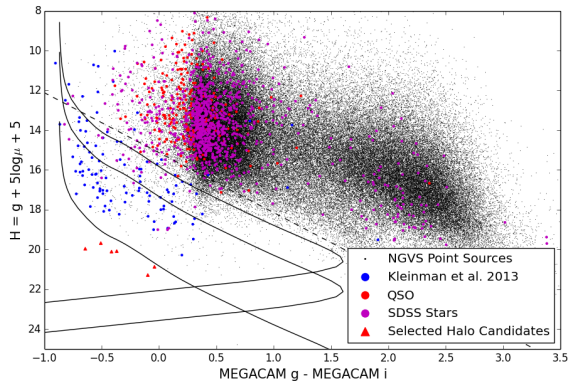}
	\caption{Reduced proper motion diagram for WD candidates. Spectroscopically confirmed WDs, QSOs, and stars are plotted as in Figure \ref{fig:modelcc}. Red triangles are halo candidates selected in this work as detailed in \S\ref{sec:Halo}. Solid lines represent tracks for model WDs with a pure hydrogen atmosphere with v$_{t}$ = 20, 40, and 200 km~s$^{-1}$ from \cite{2006AJ....132.1221H}.}
	\label{fig:rpmd}
\end{figure}

The RPMD is a useful tool for separating WDs from main-sequence stars due to their faint absolute luminosities \citep{2001Sci...292..698O,2003ApJ...586L..95V,2006AJ....131..582K,2011MNRAS.417...93R,2015ApJS..219...19L,2017AJ....153...10M}. The selection of WD candidates can be made by combining the observed magnitudes, colors, and proper motions, and comparing them to model absolute magnitudes and tangential velocities. Model absolute magnitudes were calculated using the color-absolute magnitude relations from \cite{2006AJ....132.1221H},\cite{2006ApJ...651L.137K},\cite{2011ApJ...730..128T}, and \cite{2011ApJ...737...28B}\footnote[1]{\tt http://www.astro.umontreal.ca/$\sim$bergeron/CoolingModels} and combining an adopted tangential velocity to separate WD candidates. This can be seen in Figure \ref{fig:rpmd}, where the black lines represent tangential velocities of 20 km~s$^{-1}$, 40 km~s$^{-1}$, and 200 km~s$^{-1}$ --- values that are representative of the thin disk, thick disk, and halo, respectively \citep{2006AJ....131..571H}. Also plotted in Figure \ref{fig:rpmd} is a dashed line defined by \cite{2006AJ....131..582K}; WDs are expected to fall below this relation. As expected, the large number of known main sequence stars and QSOs in our sample are found above the dashed line. 

Using a tangential velocity cut of $v_t = $ 30 km~s$^{-1}$, we find a RPMD-selected sample of 342 WD candidates. 

\subsubsection{Common Proper Motion Pair}

The selection method also yielded one common proper motion pair, SDSS J122319.19+050121.4 and SDSS J122319.57+050121.3. These objects can be important for the study of type Ia supernovae progenitors, and also allow for the study of mass loss during the late stages of stellar evolution by constraining the IFMR \citep[e.g.,][]{1997ApJ...489L..79F}.

Visual WD binaries are rather rare, and represent roughly 5-6\% of the total number of binary pairs \citep{1991AJ....101.1476S}. The SDSS has the largest sample of wide double degenerate binaries to date. \cite{2014MNRAS.440.3184B} selected 53 candidate double degenerate pairs within SDSS DR7.
	
The WDs in our observed pair have an estimated distance of 153$\pm$5 pc, a separation of 5\farcs8, and a tangential velocity of 67$\pm$2~km~s$^{-1}$ --- consistent with being a member of the thin disk. An SDSS spectrum is also available for J122319.57+050121.3 and confirms it to be a WD. Other calculated parameters are presented in Table \ref{table:cpm}, and show that the two WDs are consistent with being a double degenerate binary.
\begin{table*}[!t]
	
	\begin{center}
		\caption{Common Proper Motion Pair Properties}
		\label{table:cpm}
		\scalebox{0.85}{
			\begin{tabular}{cccccclccc}
				\hline \hline
				SDSS ID &  $u$ (AB mag) & $g$ (AB mag)& $i$ (AB mag) & $z$ (AB mag)   & $T_{\rm eff}$~(K)& d~(pc)& $v_{t}$ (km~s$^{-1})$&$\mu_{RA} $(mas/yr)&$\mu_{DEC}$(mas/yr)\\    	
				\hline
				J122319.19+050121.4&18.46$\pm$0.02&18.00$\pm$0.01&18.25$\pm$0.01&18.53$\pm$0.04&10500$\pm$1000&153$_{-3}^{+5}$&67.1$\pm$2.3&-91.4$\pm$0.4&12.9$\pm$0.4\\
				J122319.57+050121.3&19.20$\pm$0.028&18.82$\pm$0.01&18.82$\pm$0.01&18.98$\pm$0.05&8500$\pm$500&152$_{-3}^{+5}$&66.6$\pm$2.3&-91.4$\pm$0.6&13.1$\pm$0.5\\
				\hline
			\end{tabular}
		}
	\end{center}
	\tablecomments{Magnitudes are from the SDSS.}
	
\end{table*}
\begin{table}[!t]
	
	\caption{Contamination Rates for WD Selection Methods}
	\label{table:contamination}
	\resizebox{\columnwidth}{!}{

		\begin{tabular}{ccccccc}
			\hline \hline
			Method &  $T_{\rm eff}$ &N$_{\rm WD}$&$v_r$&     QSOs       & Stars   & $f_c$ \\    
			& (K)&& matches && & (\%)\\  
			\hline
			NGVS   &    $\gtrsim$ 12\,250     & 1209&  86   &    3  &   9   & 14 \\
			
			NGVS-GUViCS   &         $\gtrsim$ 9500   & 856  &   77      &  9  &  4    &17\\
			NGVS-SDSS   &         All     & 342  &   76    &     1  &   2   & 4 \\
			
			\hline
			
		\end{tabular}
		
	}
	\linebreak
	\tablecomments{$T_{\rm eff}$ range determined from ($g - i$) color selection. 
		\bigskip}

\end{table}

\smallskip
\subsection{Sources of Contamination}
\label{sec:contaminates}
\bigskip

As Figures \ref{fig:NGVScc}-- \ref{fig:rpmd} show, some objects that are {\it not} WDs can still fall within our adopted selection regions. The primary source of contamination at very blue colors is O-subdwarf (sdO) stars. These stars are thought to be the cores of red giants that ejected their surrounding shell prior to reaching the AGB phase \citep{2009ARA&A..47..211H}. At redder colors, the most common contaminants are QSOs with emission lines that fall in the $u^*$ and $g$ bands.

The contamination rate for each method was estimated by matching the WD candidates to a compendium of spectroscopic redshift measurements available for objects in the NGVS footprint. This NGVS spectroscopic catalog includes redshifts in SDSS DR12, NED, and numerous NGVS programs that targeted the Virgo cluster \citep[for more details, see][]{2014ApJ...797..102R}. 

\begin{figure*}[!t]
	\centering
	\includegraphics[angle=0,width=0.99\textwidth]{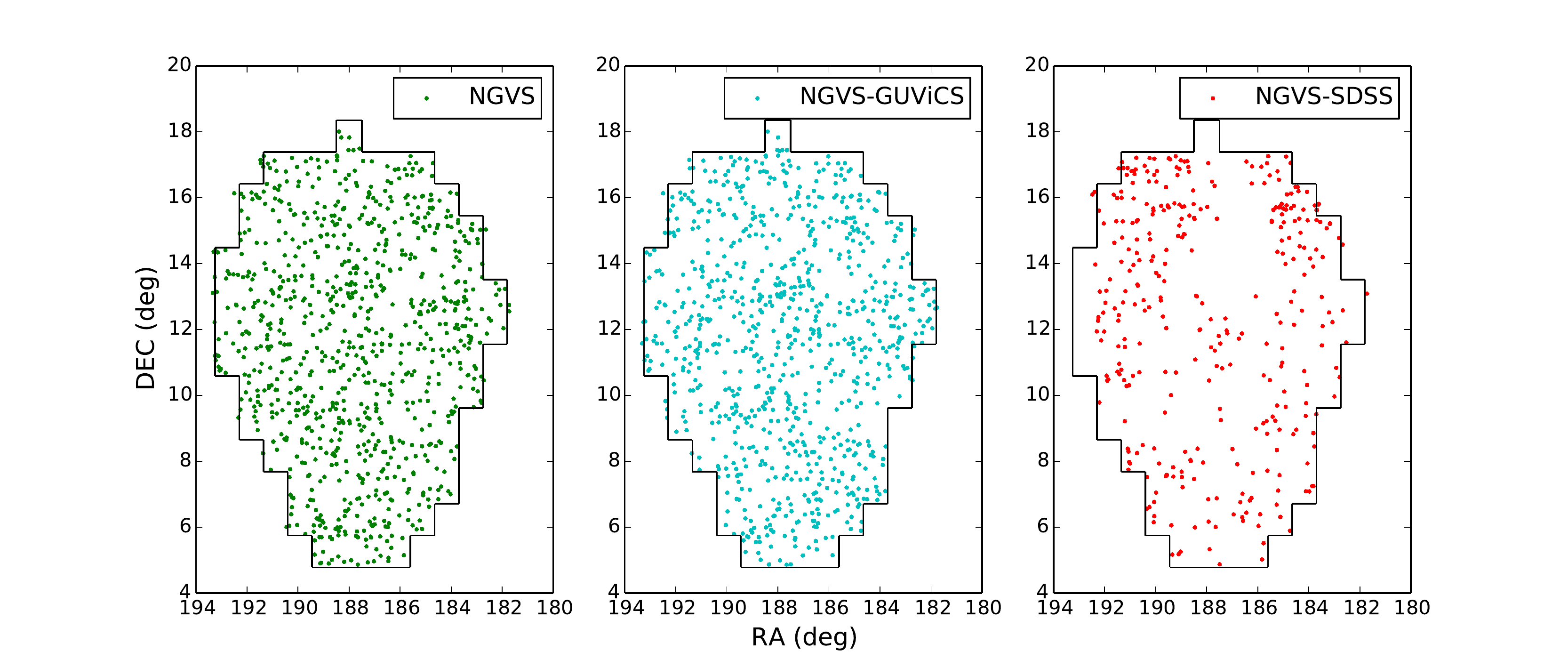}
	\bigskip
	
	\caption{WD candidates from the NGVS (left), NGVS-GUViCS (middle), and NGVS-SDSS (right) selection methods are plotted on the sky. The black lines show the approximate boundaries of the NGVS. }
	\label{fig:sky_dist}
	\bigskip
	
\end{figure*}

Table \ref{table:contamination} summarizes the results of matching the WD candidates selected by each method to the NGVS spectroscopic catalog. From left to right, the columns of this table record the selection technique, the temperature range probed by each method (see \S4.3), the number of WD candidates, the number of WDs matched to the NGVS spectroscopic catalog, the number of spectroscopically confirmed QSOs and main-sequence stars, and the overall contamination rate, in percent, for the subset of WD candidates having measured radial velocities. After inspecting the spectra of these objects, we conclude that one likely WD was misclassified as a brown dwarf, and three more likely WDs were classified as B-type stars from spectra with low signal-to-noise (i.e., S/N $\sim$ 3). The resulting contamination rate is 14-17\% for the photometrically selected samples and drops to 4\% when proper motions are included. However, as these values are dependent on the completeness of the SDSS spectroscopic catalogs they should be treated as lower limits and are meant to show the relative contamination rates between the selection methods.

\section{Photometric Properties}
\label{sec:Properties}

This section discusses photometric properties derived for the three WD catalogs described in \S3. The distribution of the WDs on the sky is presented in Figure \ref{fig:sky_dist}. While the NGVS distribution (left) is rather uniform, the voids in the NGVS-GUViCS sample (center) represent the location of bright stars that were avoided during data acquisition (see \S2.2). Furthermore, the NGVS-SDSS catalog (right) shows the location of the fields for which the stacked images were composed of exposures taken over more than one observing run (see \S2.3).

\subsection{Distributions of Apparent Magnitude}

\begin{figure}[!t]
	\centering
	\includegraphics[angle=0,width=0.49\textwidth]{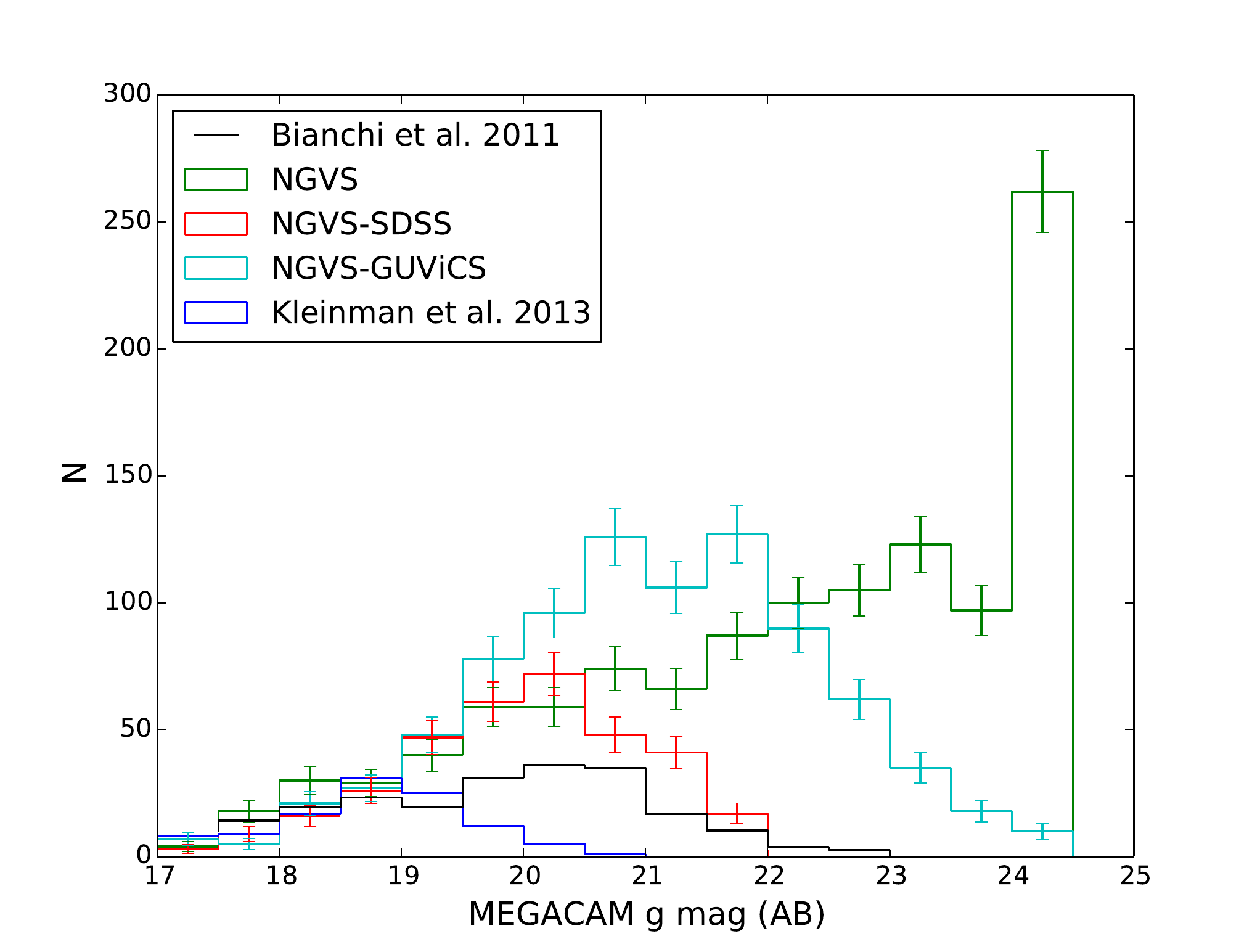}
	\bigskip
	
	\caption{Magnitude distributions for WD candidates selected using our three methods described in \S\ref{sec:selection} (green, cyan and red histograms). For comparison, we show the distributions obtained for the NGVS footprint using the SDSS spectroscopic catalog of \cite{2013ApJS..204....5K}, and GALEX AIS photometric selection from \cite{2011MNRAS.411.2770B}. The latter have been scaled to correct for incompleteness.}
	\label{fig:mag_dist}
	\bigskip
	
\end{figure}

The distribution of $g$-band magnitudes for the WD samples obtained using our three different methods is shown in Figure \ref{fig:mag_dist} (green, red, and cyan histograms). For comparison, the black histogram shows the distribution found when using the subsample of WDs from \cite{2011MNRAS.411.2770B} in the NGVS footprint; recall that this sample is based on a GALEX/AIS-SDSS color-color selection. Finally, we show (as the blue histogram) the sample of spectroscopically confirmed WDs from SDSS in the NGVS footprint from \cite{2013ApJS..204....5K}.

 The NGVS-selected LFs shown in Figure \ref{fig:mag_dist} reach to $g \sim$ 22-24.5, with the precise limit depending on the selection method. We emphasize that this is much brighter than the NGVS detection limits. Source completeness in the NGVS has been discussed in (\citeauthor{2012ApJS..200....4F} \citeyear{2012ApJS..200....4F} and Ferrarese et~al. 2017, in press), but Figure~7 of the former paper demonstrates that the 10$\sigma$ detection limit for point sources in the NGVS is $g \simeq$ 26, or 1.5--4 mag fainter than the samples considered here (which were truncated at $g$ = 24-25 mag in order to minimize contamination by background galaxies; see Figure~1). Thus, our WD samples are essentially 100\% complete to our adopted limits.

Figure \ref{fig:mag_dist} highlights the depth of the combined NGVS and GUViCS data. All five histograms agree quite well brighter $g \sim 19$; however, fainter than this, the SDSS spectroscopy and GALEX/AIS-SDSS samples clearly suffer from incompleteness. The three NGVS samples described in the previous section all peak well below this magnitude, and the sample of WDs selected entirely from the NGVS continues to rise down to the magnitude limit imposed in \S\ref{sect:data}. By contrast,
the NGVS-GUViCS sample shows a gradual fall-off beginning at $g \sim 22$ --- a result of the GUViCS  survey limit.  Similarly, the decline in the number of NGVS-SDSS candidates is an artifact of the SDSS completeness limit. 

A noteworthy feature of Figure \ref{fig:mag_dist} is the large number of objects in faintest bin of the NGVS-selected sample. As Figure \ref{fig:concentration_index} shows, the measured concentration indices for objects fainter than $g \sim 24$ are much noisier than those for brighter objects. This suggests that background sources may dominate the NGVS-selected sample in the faintest bin. For the remainder of this paper, we will discuss our findings with, and without, this faint bin.

\subsection{Photometric Distances}
\label{sec:distances}

\bigskip

Distances to a subset of the candidates were estimated using theoretical color-absolute magnitude relations from \cite{2006AJ....132.1221H}. Absolute magnitudes were estimated using a 0.6$M_{\odot}$ pure hydrogen atmosphere model \citep{2006AJ....132.1221H} and were combined with the apparent magnitude to derive a distance,
\begin{equation}
\begin{array}{lcl}
d & = & 10^{ 0.2(m-M+5-A_{g})}. \\
\end{array}
\end{equation}
Here $m$ refers to the $g$-band magnitude, $M$ is the theoretical absolute magnitude from the model of \cite{2006AJ....132.1221H}, and $A_{g} = 0.066$~mag is the average $g$-band extinction coefficient in the NGVS field.

\begin{figure}[!t]
	
	\includegraphics[angle=0,width=0.49\textwidth]{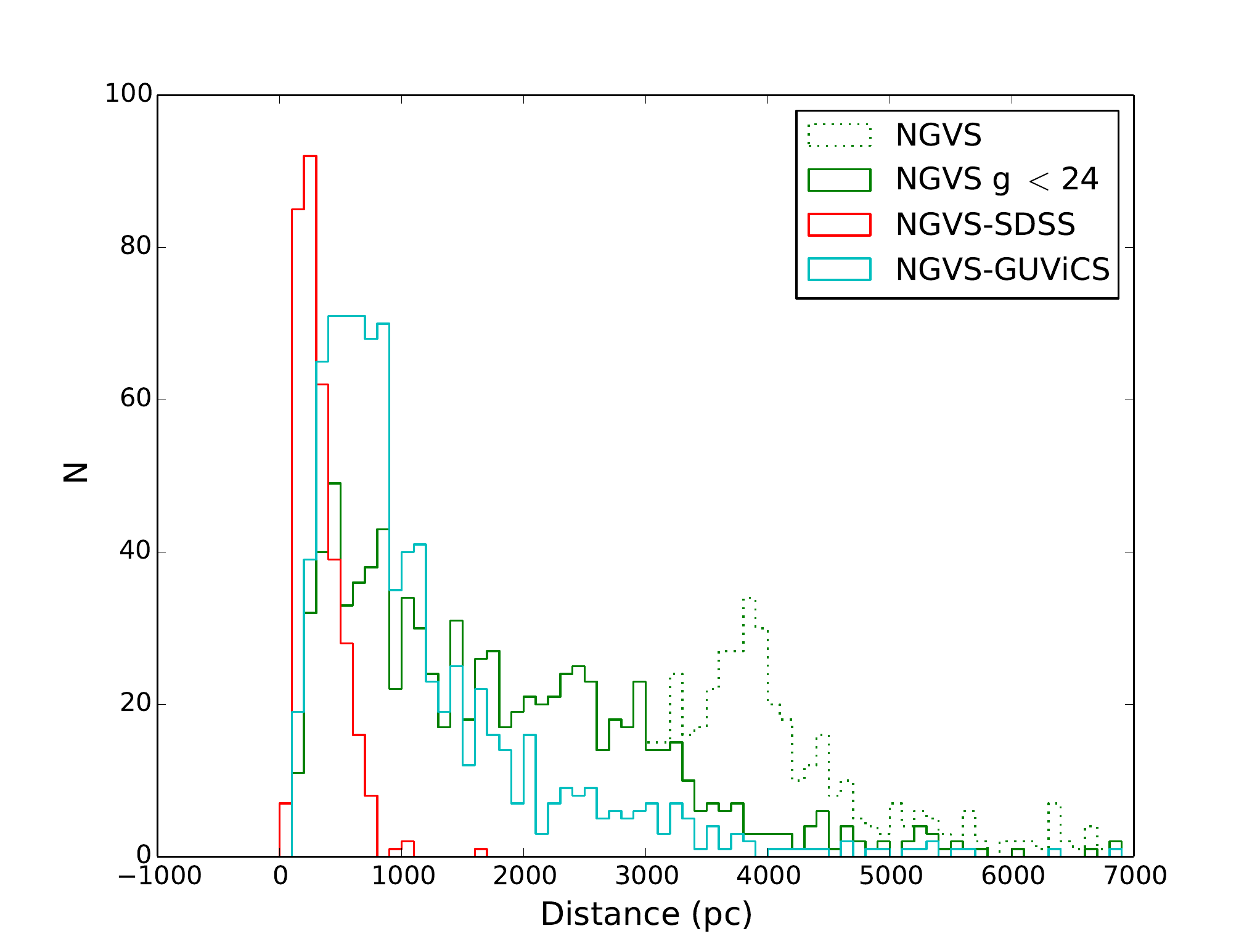}
	
	\caption{  Photometric distances for the WD candidates selected as described in \S\ref{sec:selection}. Absolute magnitudes have been calculated using the \cite{2006AJ....132.1221H} color-absolute magnitude relationship. The NGVS sample (green) is plotted with (dotted) and without (solid) the g=24.5 mag bin. 
		\bigskip}
	\label{fig:distances}

\end{figure}

The distributions in Figure \ref{fig:distances} can be explained by a combination of the survey parameters and selection methods. For instance, the NGVS-selected sample probes hotter candidates to fainter $g$-band magnitudes, and hence larger distances, than the NGVS-GUViCS sample. The NGVS-SDSS proper motion method is sensitive to objects with large proper motions, which results in the preferential selection of nearby objects.

It is important to note that the model color-absolute magnitude relation becomes very steep at high temperatures. For blue objects, this leads to large uncertainties in the derived absolute magnitudes and distances: i.e., a change in ($g-i$) color of just 0.01 mag can result in a 1 mag change in absolute magnitude. These errors can lead to a flattening of the distance distribution, yielding some objects with significantly overestimated distances. 

Another feature of Figure \ref{fig:distances} is the large number of objects in the dotted green histogram that lie at 4 kpc. This histogram represents the faintest magnitude bin (24.0 $<$ g $<$ 24.5) from Figure \ref{fig:mag_dist} and, as previously discussed, many of these objects are thought to be contaminants. We therefore believe that this feature is probably not real.

\begin{figure*}[!t]%
	\includegraphics[angle=0,width=.5\textwidth]{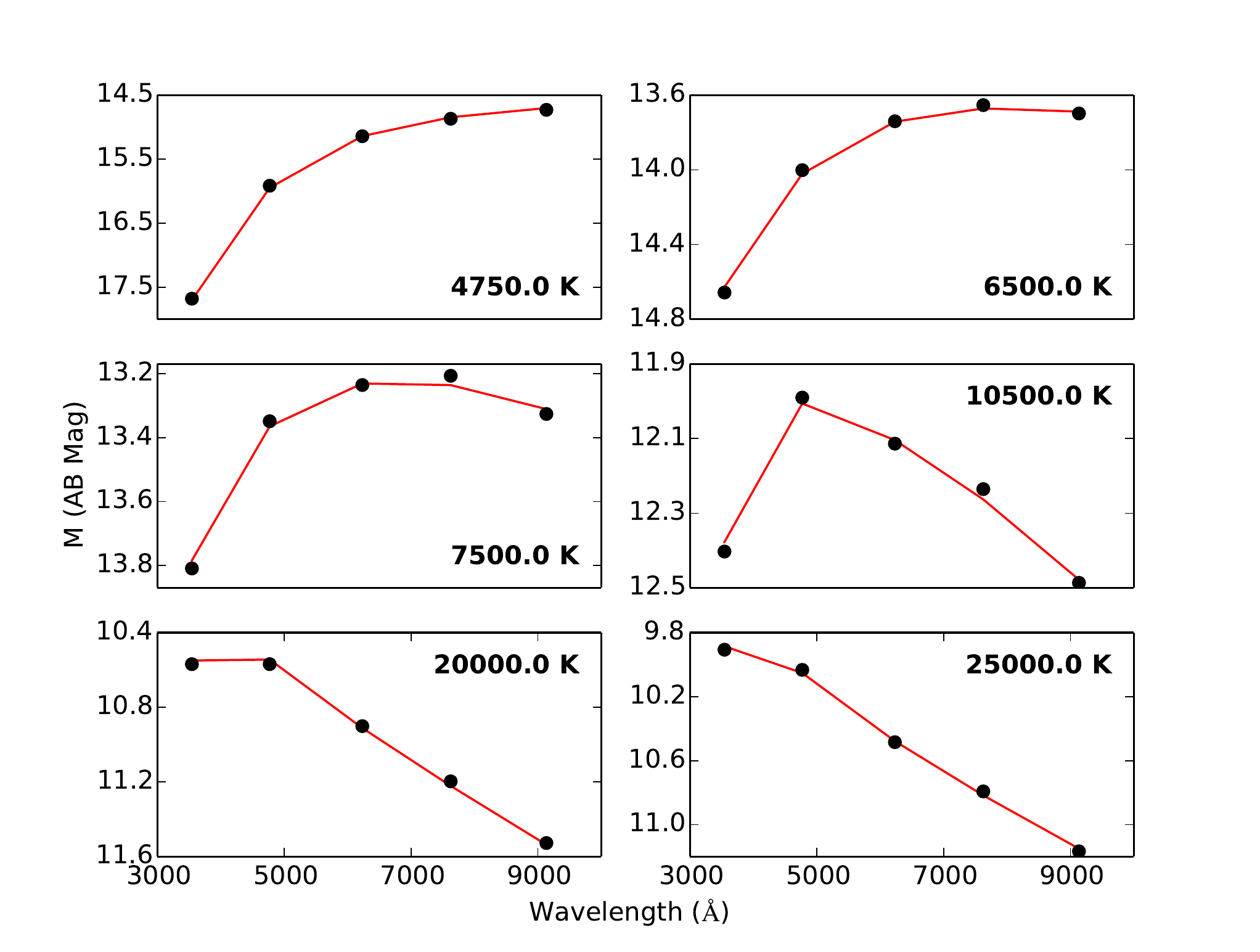}%
	\includegraphics[angle=0,width=.5\textwidth]{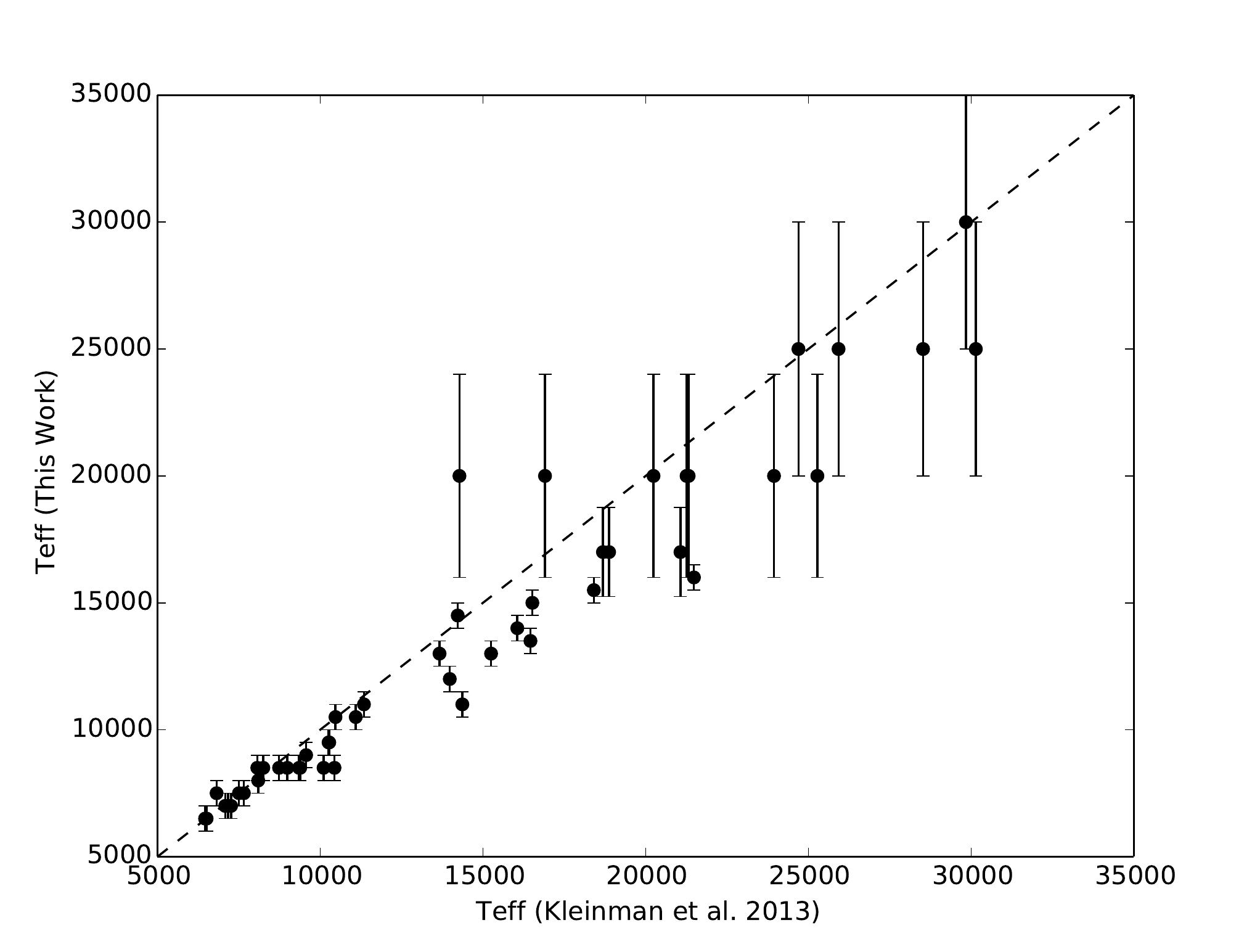}%
	
	\caption{Right: Example SED fits used to derive temperatures for WD candidates. The red lines are the best-fitting 0.6 M$_{\odot}$ SEDs for a pure hydrogen  from \cite{2006AJ....132.1221H} and the black points are the observed magnitudes after applying a vertical shift. Left: Comparison between the resulting temperatures (y-axis) and those derived from spectroscopy by \cite{2013ApJS..204....5K}. }
	\label{fig:Teff_derive}
\end{figure*}

\subsection{Effective Temperatures}    

Effective temperatures for the WD candidates were derived by fitting model spectral energy distributions (SEDs) to the $u^*giz$ photometry as in \cite{2013ApJ...766...46H}. We used the cooling models of \cite{2006AJ....132.1221H} which span the temperature range 100\,000 to 1500~K. This broad range is sampled in increments of: (1) 5000~K between 100\,000 and 20\,000~K; (2) 500~K between 17\,000 and 5500~K; and (3) 250~K between 5500 and 1500~K. Best-fit effective temperatures were computed via $\chi^2$ minimization after applying a vertical shift equal to the mean difference between the model and apparent magnitudes. The left panel of figure \ref{fig:Teff_derive} shows an example of this derivation for six candidates over a wide temperature range. A comparison between the temperatures derived via this method and those derived from SDSS spectroscopy by \cite{2013ApJS..204....5K} is presented in the right panel of \ref{fig:Teff_derive}. Overall, our method favors cooler temperatures than those measured spectroscopically. However, this effect is likely a result of the fact that both methods use different model atmospheres.

\begin{figure}[!t]%
	\includegraphics[angle=0,width=.5\textwidth]{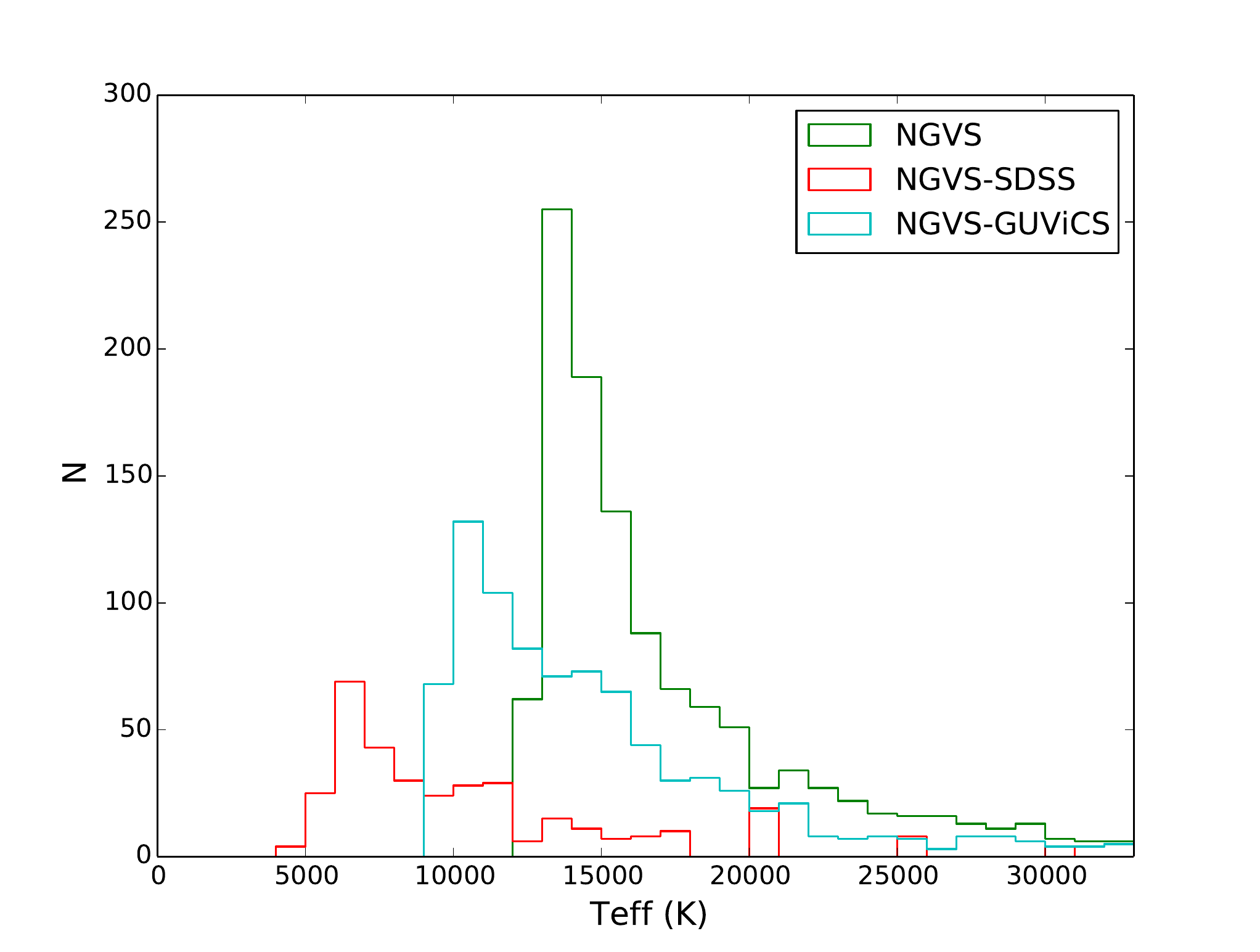}%
	\caption{Temperature distributions for WD candidates based on SED fitting.
		\bigskip}
	\label{fig:Teff}
\end{figure}

The distributions of best-fit temperatures for our three WD samples are shown in Figure~\ref{fig:Teff}. 
The NGVS-selected sample consists of 1209 WD candidates with ($g - i$) $< -0.4$, corresponding to temperatures $T_{\rm eff} \gtrsim $ 12\,500 K. By contrast, the ($g - i$) $< -0.15$ selection adopted for the NGVS-GUViCS sample (856 objects) corresponds to temperatures $T_{\rm eff} \gtrsim$ 9500~K. The sample of 342 WD candidates selected from NGVS-SDSS proper motions selection imposed no color cut, so there is no restriction on temperature in this case.

\bigskip

\begin{figure}[!t]%
	\includegraphics[angle=0,width=.5\textwidth]{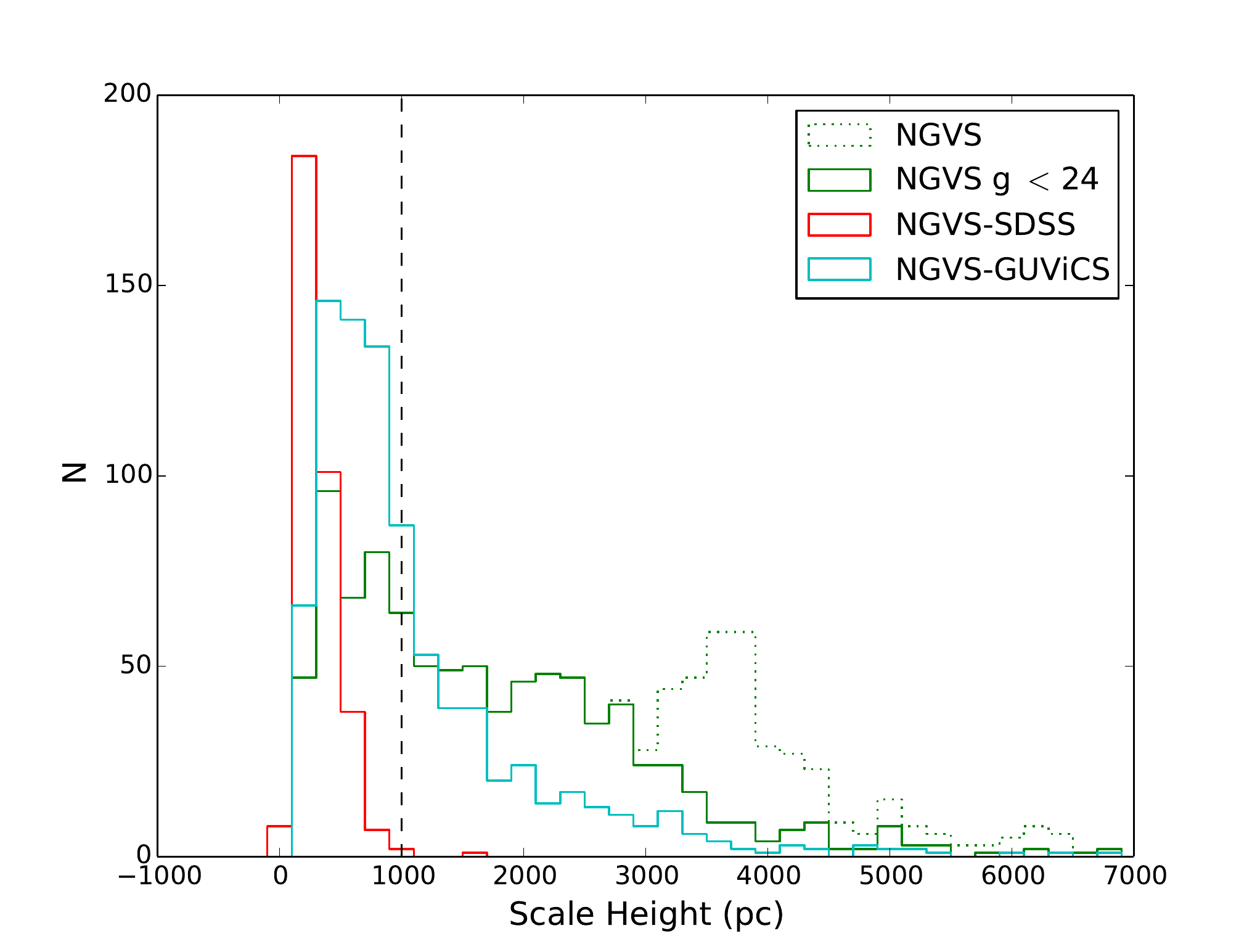}%
	\caption{Calculated scale heights based on distance estimates from the \cite{2006AJ....132.1221H} model.}
	\label{fig:scaleheight}
	\bigskip
\end{figure}

\section{Characterizing the White Dwarf Candidates}
\label{sec:Halo}

In this section, we describe our efforts to identify disk and halo WD candidates using their scale heights, proper motions, and space velocities.

\smallskip

\subsection{Photometric Samples}
\bigskip

The numbers of halo WDs in the NGVS and the NGVS-GUViCS samples were estimated using distances derived from the models of \cite{2006AJ....132.1221H}. \cite{2012ApJ...751..131B} have recently used G dwarfs as tracers of the stellar disk and found that the population is well fitted by an exponential up to a scale height of $\sim$~1 kpc. 

The calculated scale heights for each sample can be seen in Figure \ref{fig:scaleheight}, where the black dotted line represents the chosen boundary between the disk and halo populations. The depth of the NGVS sample results in the largest fraction of halo WDs, with 64\% of objects having an estimated scale height in excess of 1 kpc. This is followed by a halo fraction of 28\% in the NGVS-GUViCS sample, and just 0.3\% in the NGVS-SDSS proper motion sample. This result reflects the selection methods discussed in \S3. The NGVS and NGVS-GUViCS samples are much deeper than the NGVS-SDSS sample and hence probe larger distances.

\begin{figure}[!t]%
	\includegraphics[angle=0,width=.5\textwidth]{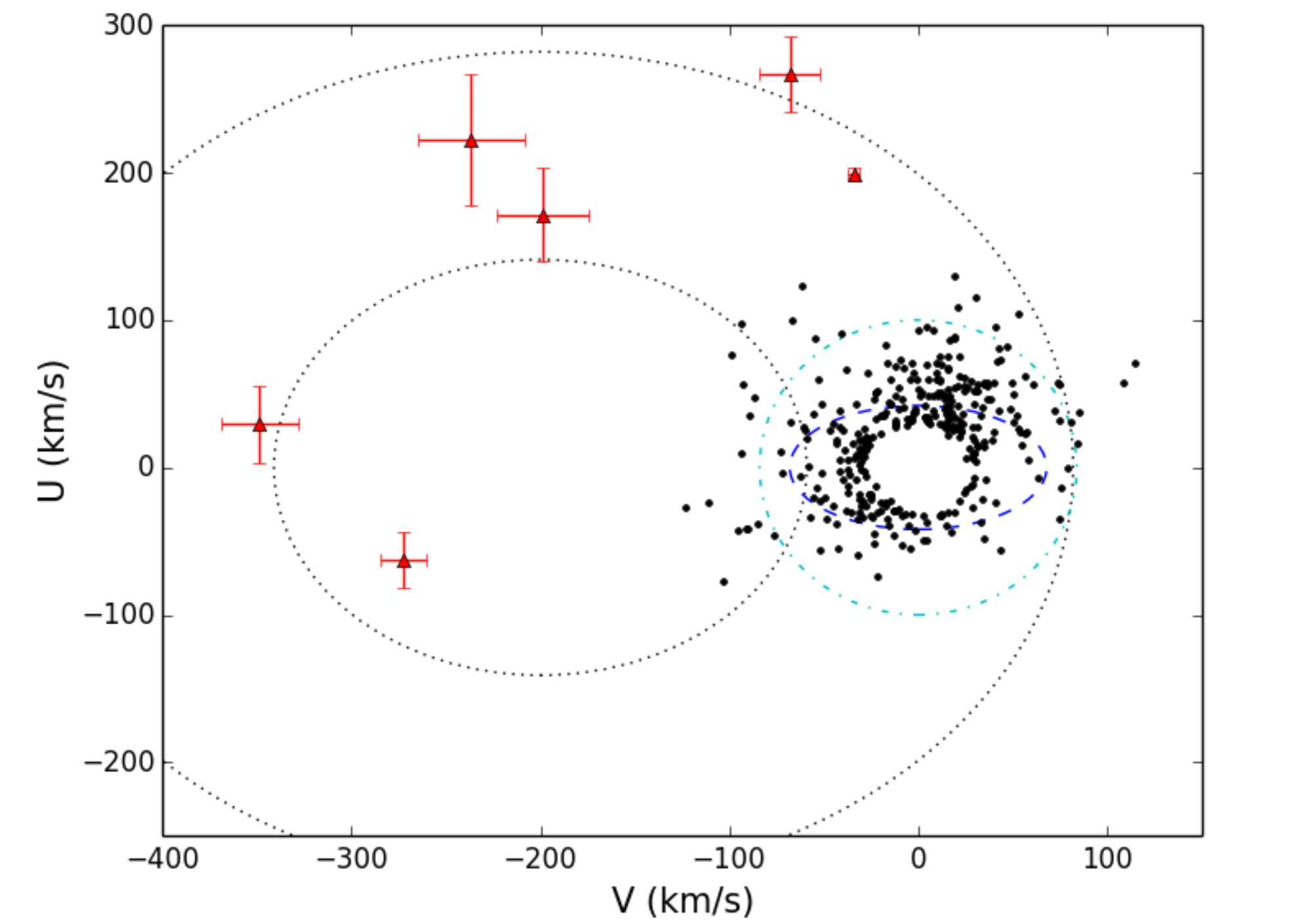}
	\caption{Galactic space velocities calculated under the assumption of zero radial velocity for NGVS-SDSS WD candidates (black). Halo candidates selected using the reduced proper motion diagram are shown as red triangles. The cyan dot-dashed and blue dashed ellipses represent the 2-$\sigma$ velocity ellipsoids for the thick and thin disk populations while the dotted black lines show the 1- and 2-$\sigma$ velocity ellipsoids for the halo, all from \cite{2000AJ....119.2843C}. }
	\label{fig:uv_planeerr}
	\bigskip
\end{figure}%

\smallskip
\subsection{Kinematic Sample}
\bigskip

\subsubsection{Reduced Proper Motion Diagram}

The RPMD can be used to characterize WDs as probable disk or the halo members based on their estimated tangential velocities \citep[e.g.,][]{2006AJ....131..582K,2011MNRAS.417...93R}. Figure \ref{fig:rpmd} shows the model curves of \cite{2006AJ....132.1221H} for 0.6$M_{\odot}$ WDs with a pure hydrogen atmosphere and tangential velocities of 20, 40, and 200 km~s$^{-1}$ --- appropriate for the thin disk, thick disk, and halo, respectively. 

A total of seven halo WD candidates lie below the $v_t = 200$ km~s$^{-1}$ relation. A visual inspection of the SDSS and NGVS imaging reveals one of the objects to be a binary that is resolved in the NGVS but unresolved in the SDSS, leading to a spurious proper motion. The properties of the remaining six halo candidates are discussed in \S\ref{sec:Discussion} and summarized in Table~\ref{table:halo_candidates}.

\subsubsection{Galactic Space Velocities}

It is possible to separate disk and halo stars using their Galactic space velocities (U, V, W) \citep[e.g.,][]{Pauli2006}. Since these velocity components are defined with respect to the Galactic center, it is necessary to transform their positions from equatorial ($\alpha$, $\delta$) to Galactic (\textit{l}, \textit{b}) coordinates \citep[e.g.,][]{1987AJ.....93..864J}.  
Needless to say, the transformation into the (U, V, W) frames requires both a proper motion and a radial velocity. Since radial velocities are not available for the vast majority of our WD candidates, we set $\rho$ to zero. However, the contribution to U and V space velocity components is small at the fairly high Galactic latitude of the NGVS ($\textit{b}\sim 75^{\circ}$).

\smallskip

With this caveat in mind, the U and V velocities for our WD candidates are shown in Figure \ref{fig:uv_planeerr}. The cyan dot-dashed and blue dashed ellipses represent the 2-$\sigma$ velocity ellipsoids for the thick and thin disk populations, respectively. The dotted black lines show the 1- and 2-$\sigma$ velocity ellipsoids for the halo. These velocity ellipsoids were taken from \cite{2000AJ....119.2843C}. 

The six halo WD candidates identified from the RPMD are shown as the red triangles in Figure~\ref{fig:uv_planeerr}. Of these six objects, five fall inside the 2-$\sigma$ velocity ellipsoid of the Galactic halo, while the sixth lies just outside. 
\begin{figure}[!t]
	\includegraphics[angle=0,width=.5\textwidth]{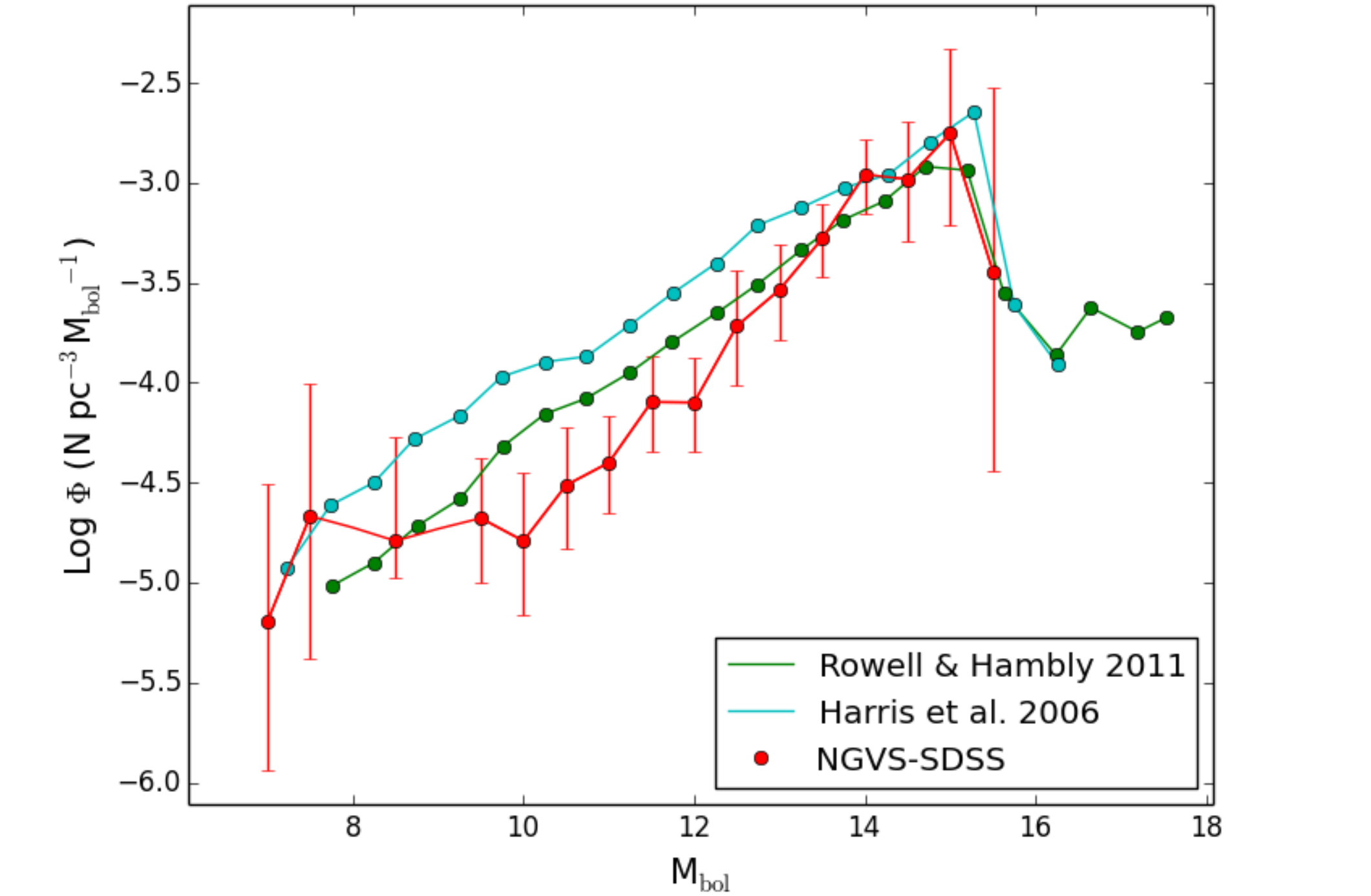}
	\caption{SDSS-NGVS luminosity function (red) compared to those derived by \cite{2006AJ....131..571H} (cyan) using SDSS DR3 and \cite{2011MNRAS.417...93R} (green) using the SuperCOSMOS survey. }
	\label{fig:previous_data}
	\bigskip
\end{figure}

\smallskip
\subsection{Sample Completeness and Luminosity Functions}
\bigskip

Sample completeness can be assessed using the V/V$_{max}$ method \citep{1968ApJ...151..393S} which gives the ratio of the volume out to a given detected WD to the volume out to the distance the WD would have at the detection limit of the survey. V$_{max}$ can be calculated using the methodology described by \cite{2011MNRAS.417...93R} and \cite{2013ApJ...766...46H}:
\begin{equation}
\begin{array}{lcl}
V_{max} & = & \beta \int_{r_{min}}^{r_{max}} \frac{\rho}{\rho_{\odot}}R^{2}dR, \\
\end{array}
\end{equation}
Here $\beta$ is the survey area as a fraction of the total sky, $\frac{\rho}{\rho_{\odot}}$ is the stellar density along the line of sight, $R$ is the distance, and $r_{min}$ and $r_{max}$ are the minimum and maximum distances for which an object would fall within the survey limits. These distances are defined as 
\begin{equation}
\begin{array}{lcl}
r_{min} & = & 10^{0.2(m_{min} - M + 5 - A_{g})} \\
\end{array}
\end{equation}
and
\begin{equation}
\begin{array}{lcl}
r_{min} & =  &10^{0.2(m_{max} - M + 5- A_{g})} \\
\end{array}
\end{equation}
where $m_{max}$ and $m_{min}$ are the faint and bright limits of the survey. The stellar density for the disk is assumed to be 
\begin{equation}
\begin{array}{lcl}
\frac{\rho}{\rho_{\odot}} & = & \exp\left[-\frac{|R\sin b + Z_{\odot}|}{H}\right], \\
\end{array}
\end{equation}
where $H$ is the scale height of the disk, $R$ is the distance to the object from the model of \cite{2006AJ....132.1221H}, $b$ is the Galactic latitude, and $Z_{\odot}$ = 20~pc is distance of the Sun above the Galactic plane \citep{2006JRASC.100..146R}.

Using this method, we find $\left \langle \frac{V}{V_{max}} \right \rangle$ $\pm 1/(12N)^{1/2}$ = 0.39 $\pm$ 0.02 for the sample of $N = 342$ kinematically selected WD candidates. For a uniform distribution, the mean value should be 0.5, suggesting some incompleteness in the sample. This incompleteness is likely a consequence of our selection method, which removes many faint objects due to the large uncertainty in their SDSS positions. 

The maximum volume, $V_{max}$, can be used to construct a luminosity function that, once integrated, yields a space density. The number density of objects, $\phi$, can be expressed as the sum of the inverse maximum volumes:
\begin{equation}
\begin{array}{lcl}
\phi & = & \sum_{i}^{N} \frac{1}{V_{max, i}}. \\
\end{array}
\end{equation}
The uncertainty in the number density, $\sigma_{\phi}$, can then be calculated using Poisson statistics \citep{2011MNRAS.417...93R}:
\begin{equation}
\begin{array}{lcl}
\sigma_{\phi}^2 & = & \sum_{i}^{N} \frac{1}{V^2_{max, i}}. \\
\end{array}
\end{equation}
 
\begin{figure}[!t]
	\includegraphics[angle=0,width=.49\textwidth]{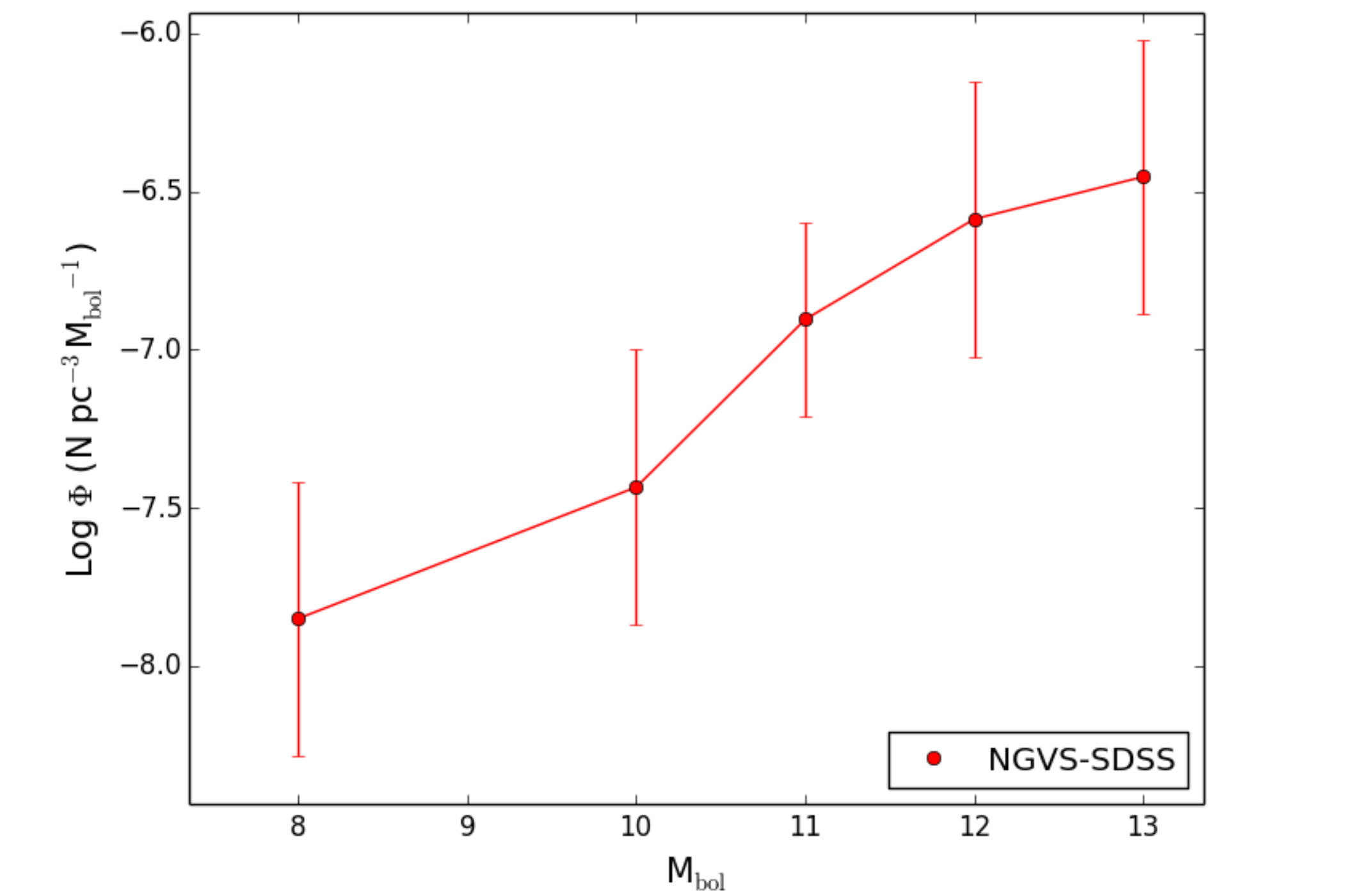}
	\caption{Luminosity function for halo candidates selected using the NGVS-SDSS proper motions. }
	\label{fig:lf_halo}
\end{figure}

The resulting luminosity function must then be rescaled to account for the absence of WDs with tangential velocities less than 30 km~s$^{-1}$ in our sample selected by proper motion. This scale factor was calculated using the Besan\c{c}on WD catalog (explained in more detail below), which suggests that the fraction of WDs having tangential velocities above 30 km~s$^{-1}$ in the NGVS field is 0.746. This value is in good agreement with the value of 0.726 obtained by \cite{2006AJ....131..571H} for the SDSS DR3 field. 
The luminosity functions for our disk and halo WD candidates are shown in Figures~\ref{fig:previous_data} and \ref{fig:lf_halo}, respectively, and are discussed below.

\section{Discussion}
\label{sec:Discussion}
\smallskip 

\subsection{Comparison to the SDSS and SuperCOSMOS Luminosity Functions}

Figure \ref{fig:previous_data} compares the luminosity function for the sample of WD candidates selected from SDSS-NGVS proper motions to those obtained with SDSS DR3 \citep{2006AJ....131..571H} and the SuperCOSMOS survey \citep{2011MNRAS.417...93R}. This comparison shows a discrepancy at the bright end of the luminosity function, although the faint ends are consistent. The discrepancy at the bright end is a likely a consequence of the high Galactic latitude of the NGVS field combined with the geometry of the Milky Way disk. That is to say, the SDSS and SuperCOSMOS surveys reached to lower Galactic latitudes, which contain different fractions of thin and thick disk stars.

\begin{figure}[!t]
	\includegraphics[angle=0,width=.49\textwidth]{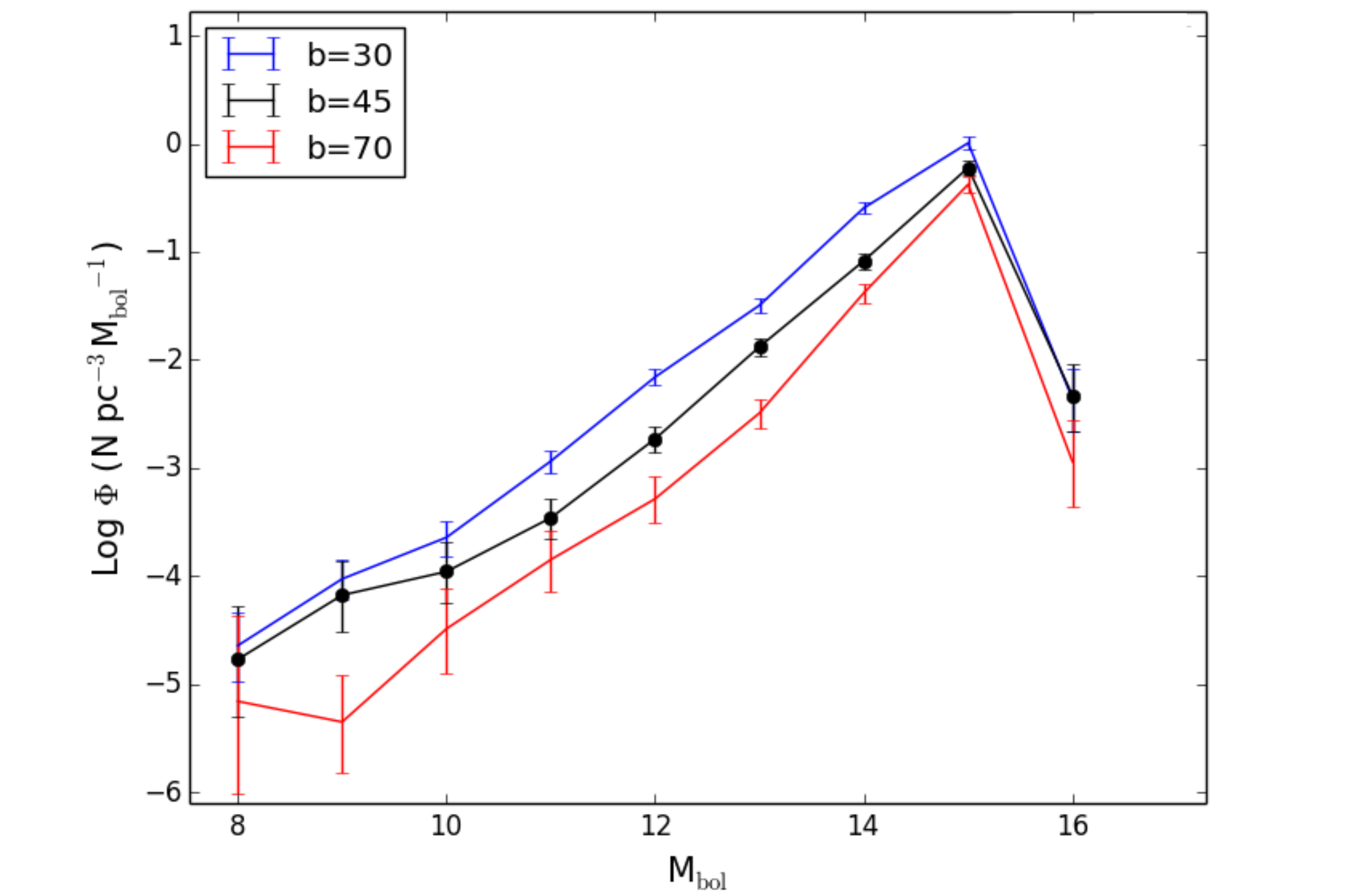}
	\caption{WD luminosity functions calculated with the Besan\c{c}on model for three different Galactic latitudes.}
	\label{fig:latitude}
	\bigskip
\end{figure}

We explored the importance of this effect using the Besan\c{c}on model. Figure \ref{fig:latitude} shows the luminosity function of WDs at $b = $ 30$^{\circ}$, 45$^{\circ}$, and 70$^{\circ}$ while keeping all other parameters of the model fixed. The luminosity functions at each Galactic latitude can be seen as the blue, black, and red curves in Figure \ref{fig:latitude}.  This exercise demonstrates that, at higher Galactic latitudes, there is a discrepancy at the bright end that is comparable to that seen in the observations, while the faint ends of the luminosity functions are nearly identical.

\subsubsection{Number Densities}

A number density for the thin disk was estimated by integrating the NGVS-SDSS luminosity function with $H = 250$ pc. Because this has been a customary choice in many previous studies, it allows for a direct comparison to earlier measurements. The resulting number density for the disk is then
\begin{equation}
\begin{array}{lcl}
\phi_d = (2.81\pm0.52) \times 10^{-3}~{\rm pc}^{-3}.\\
\end{array}
\end{equation}
This value is consistent with several previous estimates: e.g., 2.36$\pm$0.27 $\times~10^{-3}$ by Hu et al. (2013) and 3.19$\pm$0.09 $\times~10^{-3}$ by Rowell \& Hambly (2011). It is somewhat lower than the estimates of 5.5$\pm$0.1 $\times 10^{-3}$, 4.6 $\times 10^{-3}$ and 3.4 $\times 10^{-3}$ from \cite{2017AJ....153...10M}, \cite{2006AJ....131..571H} and \cite{1998ApJ...497..294L}, respectively, although this discrepancy is likely caused by the differing sightlines of the surveys, as explained above.

\begin{table*}[!t]
	
	\begin{center}
		\caption{Properties of NGVS Halo White Dwarf Candidates}
		\label{table:halo_candidates}
		\scalebox{0.85}{
			\begin{tabular}{cccccclccc}
				\hline \hline
				NGVS ID &  $u^*$ (AB mag) & $g$ (AB mag)& $i$ (AB mag) & $z$ (AB mag)   & $T_{\rm eff}$~(K)& d~(pc)& $v_{t}$ (km~s$^{-1})$&$\mu_{RA} $(mas/yr)&$\mu_{DEC}$(mas/yr)\\

				\hline
				J124516.62+170505.1&21.558$\pm$0.004&20.606$\pm$0.002&20.645$\pm$0.004&20.823$\pm$0.009&6500$\pm$500&406$_{-3}^{+4}$&218$\pm$14&-18.7$\pm$5.0&-99.1$\pm$5.3\\
				J121933.27+163829.4&21.724$\pm$0.004&21.277$\pm$0.003&21.375$\pm$0.006&21.501$\pm$0.017&7500$\pm$500&613$_{-9}^{+10}$&293$\pm$35&-94.7$\pm$3.0&-62.4$\pm$2.9\\
				J121955.46+151523.1&21.463$\pm$0.003&20.975$\pm$0.002&21.343$\pm$0.006&21.582$\pm$0.016&11000$\pm$1000&569$_{-265}^{+267}$&201$\pm$96&-59.2$\pm$2.7&-30.4$\pm$2.2\\
				J122515.80+064849.2&20.924$\pm$0.003&20.661$\pm$0.002&21.075$\pm$0.005&21.324$\pm$0.014&10000$\pm$1000&758$_{-6}^{+7}$&279$\pm$26&-71.3$\pm$3.4&-31.0$\pm$3.1\\
				J123614.18+061135.0&20.547$\pm$0.002&20.204$\pm$0.002&20.711$\pm$0.004&20.941$\pm$0.014&9500$\pm$500&714$_{-7}^{+8}$&267$\pm$21&-75.2$\pm$4.8&-53.9$\pm$4.6\\
				J121455.93+125058.4&20.028$\pm$0.002&19.873$\pm$0.001&20.515$\pm$0.004&20.830$\pm$0.008&20000$\pm$3000&853$_{-11}^{+12}$&423$\pm$98&44.3$\pm$3.4&-94.7$\pm$3.4\\
				\hline
			\end{tabular}
		}
	\end{center}
	\tablecomments{WD properties calculated for an assumed mass of 0.6$M_{\odot}$.}
	\bigskip
\end{table*}

Assuming that the six high-velocity WDs belong to the halo (\S5.2.2), the number density of halo WDs is estimated to be
\begin{equation}
\begin{array}{lcl}
\phi_h = (7.85\pm4.55) \times 10^{-6}~{\rm pc}^{-3}.\\
\end{array}
\end{equation}
This value is marginally lower than the number densities of $\sim$~4~$\times~10^{-5}$~pc$^{-3}$ found by \cite{2006AJ....131..571H} and 3.5$\pm$0.7$\times~10^{-5}$~pc$^{-3}$ found by \cite{2017AJ....153...10M}. This is likely a result of our selection method, which rejected many faint objects with large proper motion errors. Our result is marginally higher than the number density of 4.4$\pm$1.3 $\times~10^{-6}$ reported by \cite{2011MNRAS.417...93R}. 

Of course, such comparisons should be treated with caution because they are highly dependent on survey parameters. The number density relies mainly on the faint end of the luminosity function, which is in turn dependent on survey depth. As the depth increases, the relative contributions from the thin disk, thick disk, and halo will also change.

\subsection{Model Comparisons}

\begin{figure*}[p]
	\centering
	\includegraphics[angle=0,width=.75\textwidth]{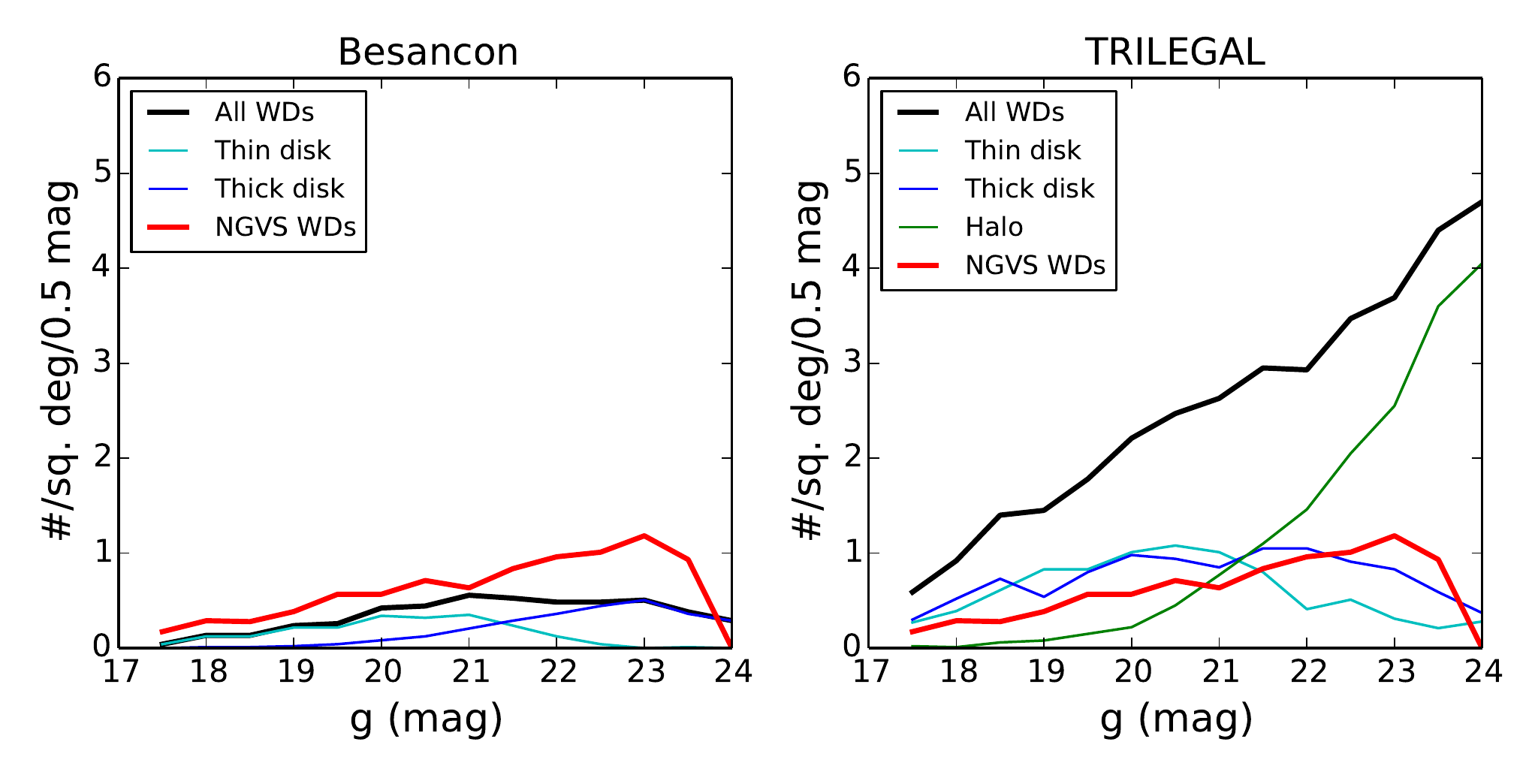}\\
	\includegraphics[angle=0,width=.75\textwidth]{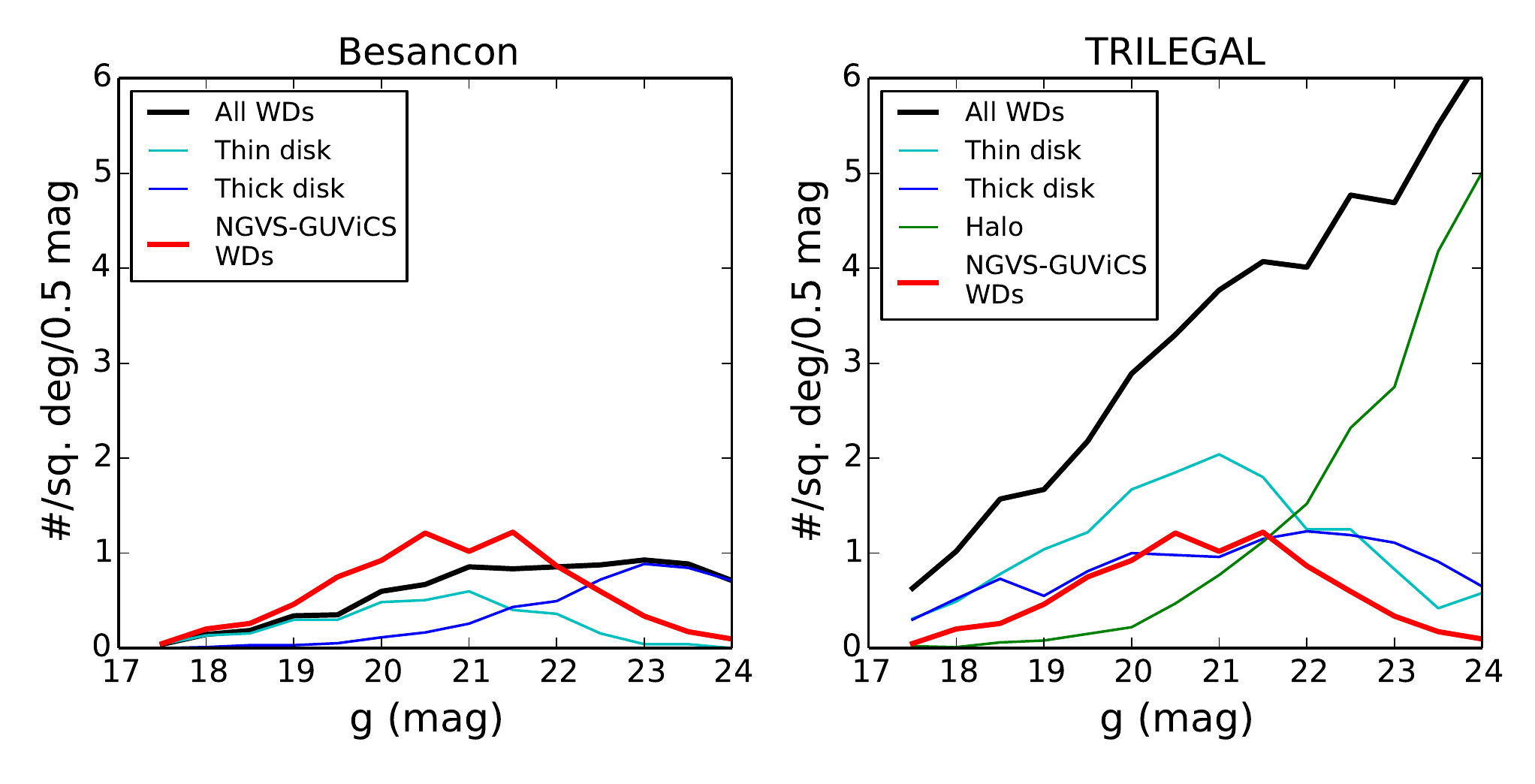}\\
	\includegraphics[angle=0,width=.75\textwidth]{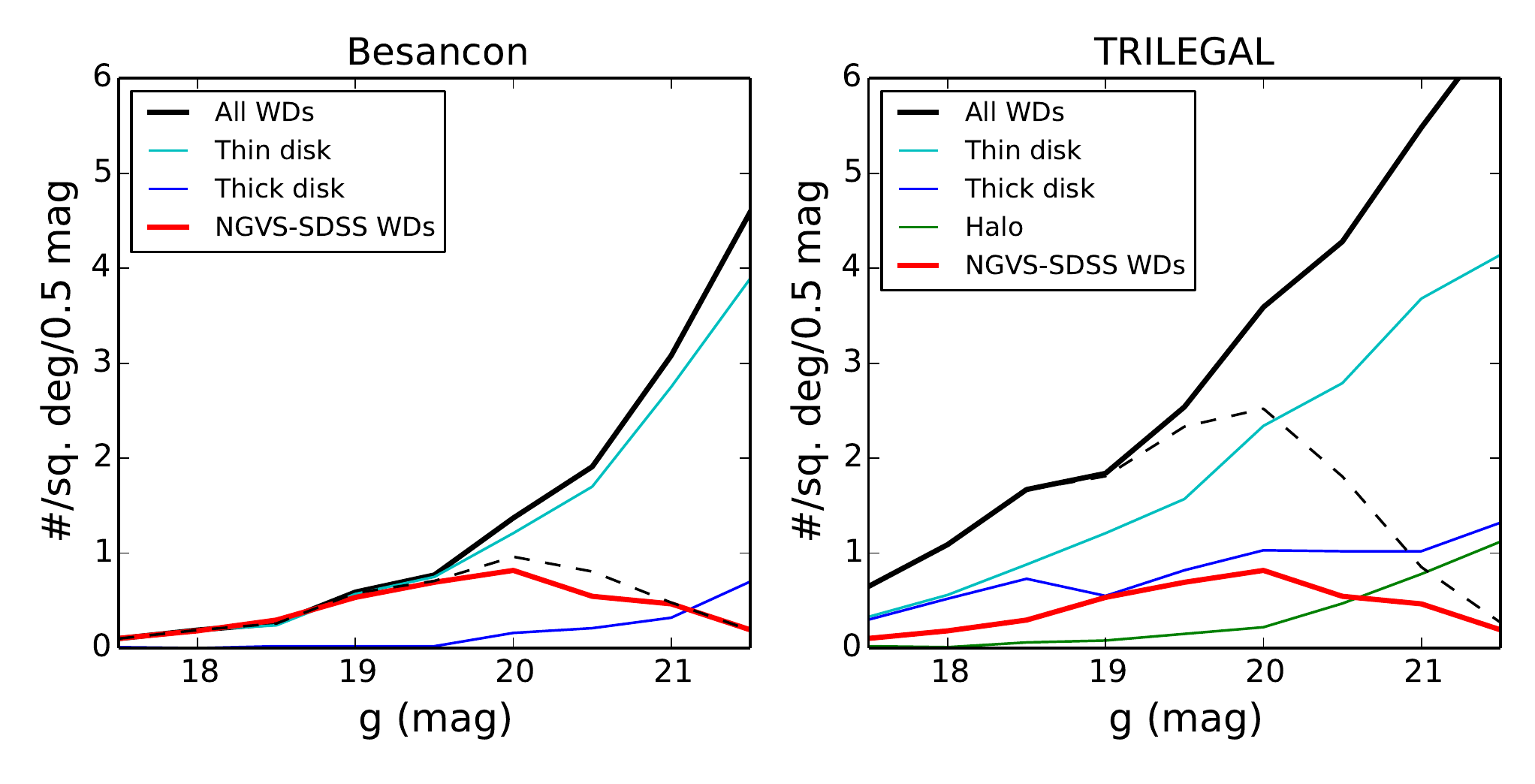}
	\caption{Magnitude distributions for the NGVS (top), NGVS-GUViCS (middle), and NGVS-SDSS (bottom) WD catalogs compared to the TRILEGAL (right) and Besan\c{c}on (left) mock catalogs. The observed WD candidates are shown in red, and the mock WDs are separated into the thin disk (cyan), thick disk (blue) and halo (green) respectively. In the lower panels, the dashed black curves show the mock WD samples after applying corrections to account for the incompleteness suffered by the data.}
	\label{fig:model_counts}
	\bigskip

\end{figure*}

In this section we compare the number densities generated by the TRILEGAL and Besan\c{c}on stellar population synthesis models with the three samples of WDs selected in Section \ref{sec:selection}.

A TRILEGAL WD catalog was computed by generating ten mock regions of 10~deg$^2$ (the maximum allowable survey area) covering the NGVS footprint. All other input parameters were left at their default values, including a Chabrier initial mass function, a Milky Way extinction model, and a three-component Galactic model, including squared hyperbolic secant thin and thick disks and an oblate halo. Due to the high Galactic latitude of the NGVS field, this sightline will include no bulge stars. Stars belonging to the mock catalog were deemed to be WDs if their log~\textit{g} $>$ 7. Mock WDs were then assigned to the appropriate Galactic component --- numbered 1 for the thin disk, 2 for the thick disk, and 3 for the halo.

A Besan\c{c}on WD catalog was constructed by generating a region of 100 deg$^2$, again centered on the NGVS footprint, with default parameters described in \cite{2003A&A...409..523R}. As with the TRILEGAL selection, WDs were identified by their high surface gravities. Objects were then divided into their respective Galactic component based on their designated population, with numbers 2-7 representing the thin disk, 8 representing the thick disk, and 9 representing the halo. 

Three catalogs were generated by each model, and WDs were selected using the same color and magnitude selections described in Section \ref{sec:selection}. The resulting g-band magnitude distributions, described as number densities to account for variations in field area, are shown in Figure~\ref{fig:model_counts}. 

The left hand panels show the comparison between the Besan\c{c}on models and the NGVS (top), NGVS-GUViCS (middle), and NGVS-SDSS (bottom) WD candidates. Using the NGVS and NGVS-GUViCS catalogs as representations for the hot, young, WD population reveals a relatively good agreement between the model and observations. The slightly larger number of observed WDs in the NGVS (top) and NGVS-GUViCS (middle) catalogs is likely due to the fact that, as discussed in \S3.4 and Table \ref{table:contamination}, both samples include a non insignificant fraction of contaminants ($\sim$15\%). Unfortunately, due to the lack of complete spectroscopic data to the limit of the NGVS survey, the contamination rate cannot be accurately modeled. Additionally, as discussed in \S4.1 and Figure \ref{fig:mag_dist}, the NGVS-GUViCS sample is known to be incomplete below g$\sim$22, which can explain the decline in observed WDs compared to the models.

Comparing the Besan\c{c}on model to the NGVS-SDSS proper motion sample reveals a strong agreement at bright magnitudes with a divergence beginning around g$\sim$19.5. This is a result of the cuts in proper motion error implemented to reduce contamination from other stellar sources. This effect has been accounted for in the black dashed line, which was calculated using the fraction of objects in each bin that passed our selection criteria and applied to the mock catalog. This shows that the model and observed counts are in agreement. 

The right hand panels of Figure \ref{fig:model_counts} compare the observed number densities to those computed using the TRILEGAL mock catalogs. The comparison between the observed densities of hot WDs from the NGVS and NGVS-GUViCS samples shows an overprediction by the TRILEGAL model, consistent with the results from \cite{2011MNRAS.411.2770B}, who showed that this is a result of the IFMR adopted by TRILEGAL. This discrepancy is also apparent in the NGVS-SDSS proper motion sample, where the selection is not based on temperature. 

Figure \ref{fig:model_counts} also highlights the disagreement between the predicted numbers of halo WDs present in our catalogs. For example, TRILEGAL predicts 280 halo WDs in our catalog selected by proper motion selected, whereas the Besan\c{c}on does not predict any. Although it predicts no halo WDs in our samples, the Besan\c{c}on model does indeed include them, but they are expected to be too faint and too cool to be detected in our survey. The discrepancy between observations and model likely arises from the method by which the mock halo WDs are generated from their main-sequence progenitors: i.e., through the IFMR, which is poorly constrained by observations. The IFMR used in the Besan\c{c}on model predicts final halo WD masses of $0.7~M_{\odot}$, which would result in fainter model magnitudes. Observations have revealed a wide range of masses for halo WDs, and favor lower masses and hence higher luminosities \citep[see, e.g][]{2011MNRAS.411.2770B}. For their sample of field halo WDs, \cite{Pauli2006} obtained masses of $0.35-0.51~M_{\odot}$, while \cite{2012MNRAS.423L.132K} calculate masses of $0.62~M_{\odot}$ and $0.77~M_{\odot}$. Recent studies of the globular cluster M4 have established that WDs with masses of $\sim0.5-0.55~M_{\odot}$ are currently being formed in halo environments  \citep[see, e.g][]{2004A&A...420..515M,2009ApJ...697..965B,2009ApJ...705..408K} and hence a model mass of $0.7~M_{\odot}$ does not accurately represent a halo environment. 

\subsection{Thick Disk or Halo Membership?}

The debate over whether halo WDs can be identified solely on the basis of their kinematics took on a renewed importance with the study performed by \cite{2001Sci...292..698O}, who used Galactic space velocities from WDs in the SuperCOSMOS Survey to conclude that as much as 2\% of the ``unseen" matter in the Galactic halo could arise from cool WDs. \cite{2003ApJ...586..201B} and \cite{2005ApJ...625..838B} emphasized the importance of total stellar ages (i.e., main-sequence plus WD cooling ages) when  characterizing WDs as belonging to the stellar halo, because many WDs with halo kinematics were so hot, and thus so young,as to call into question their association with the halo. On the other hand, if these WDs had lower than expected masses then they could have ages consistent with a halo population,  because they would have formed from less massive progenitors --- objects with longer main-sequence lifetimes, and hence, longer main-sequence and WD cooling lifetimes. For example, a 0.6$M_{\odot}$ WD would have an initial main-sequence mass of $\sim$ 2 M$_{\odot}$ and a lifetime of $\sim$ 1 Gyr. By contrast, a 0.53$M_{\odot}$ WD would have a main-sequence mass of 1 M$_{\odot}$ and a lifetime approaching 10 Gyr \citep{2016MNRAS.463.2453D}. This simple example highlights the importance of accurate mass and distance measurements when assigning WDs to different components of the Galaxy.

\cite{2001A&A...378L..53R} used the Besan\c{c}on model to show that the sample of \cite{2001Sci...292..698O} is likely dominated by the thick disk as opposed to the halo. Using our Besan\c{c}on mock catalog we compute the reduced proper motion for the mock WDs to determine the expected number of thick disk objects that would be selected as halo candidates in our work. This exercise reveals an expectation value of 1 WD, showing that the thick disk alone probably does not account for our observed sample of high velocity WDs.

The properties of the six halo WD candidates selected from our proper motions are summarized in Table~\ref{table:halo_candidates}. Temperature estimates were obtained by SED fitting  as described in \S\ref{sec:Properties}. Based on the derived temperatures, and assuming a 0.6$M_{\odot}$ model with a pure hydrogen atmosphere, the inferred cooling ages range from 60 Myr to 6 Gyr. However, as previously noted, these values are highly sensitive to the adopted mass.

\subsection{Comparison to Open and Globular Clusters}

 In environments where the WD luminosity function can be observed to faint magnitudes, WD ages can used to intercompare the star formation histories of different stellar systems or Galactic components (e.g., \citealt{2013Natur.500...51H},\citealt{2017arXiv170206984K}). Because star clusters are typically formed during a single burst of star formation, the location of the peak of the WD luminosity function provides an indicator of cluster age: i.e., the WD luminosity function in older stellar populations will peak at fainter magnitudes for the simple reason that the WDs have had more time to cool \citep{2009ApJ...697..965B}. 

\begin{figure}[!t]
	\includegraphics[angle=0,width=.49\textwidth]{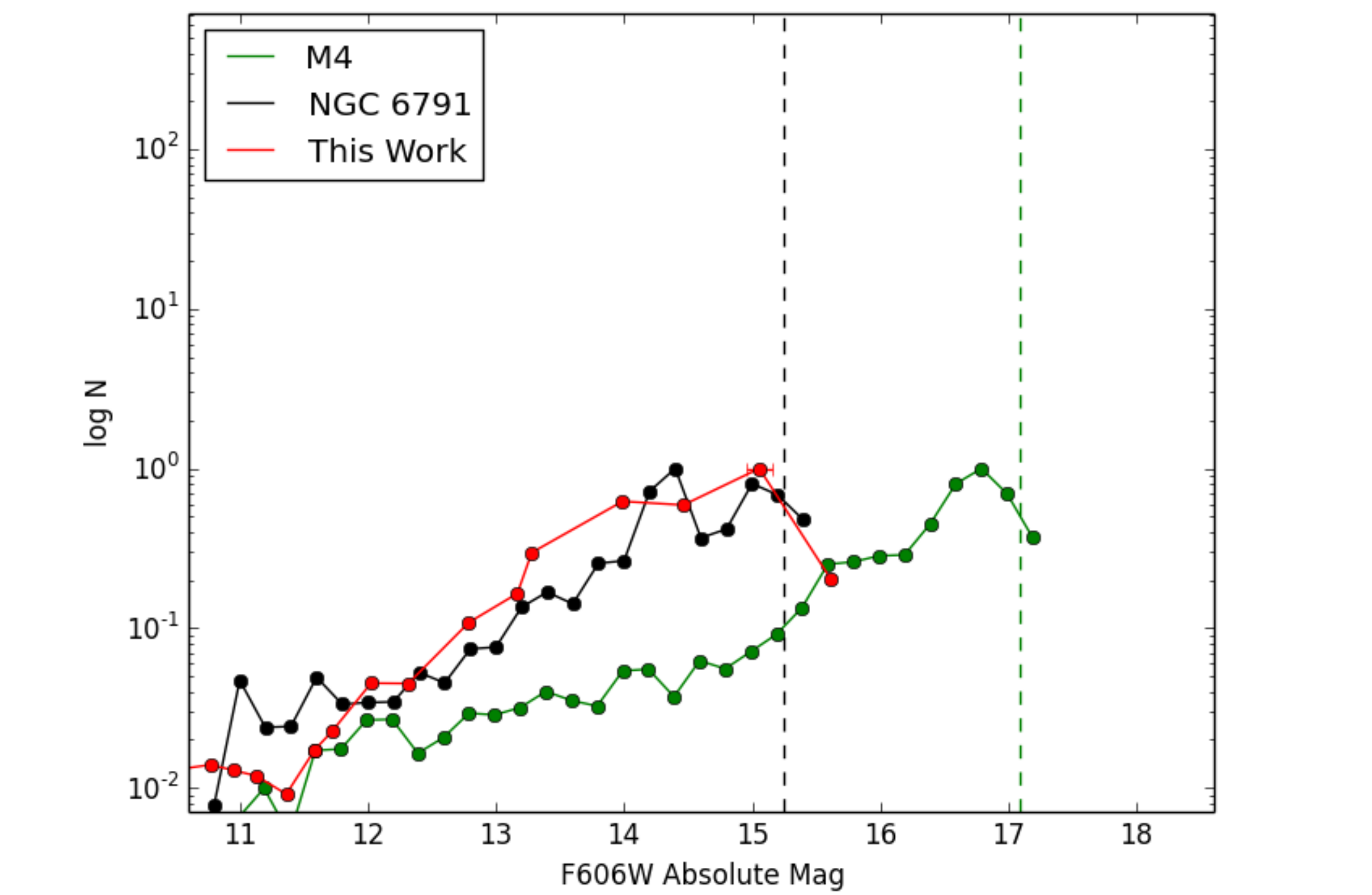}
	\caption{Luminosity function for the disk WDs compared to the open cluster NGC 6791 \citep{2008ApJ...678.1279B} and the globular cluster M4 \citep{2009ApJ...697..965B}. The 50\% completeness levels for the two clusters are indicated by the dashed lines, showing that the cluster surveys are complete below the peak of the luminosity function. }
	\label{fig:gc_lf}
\end{figure}

The stellar populations of globular and open clusters are often used as analogs for those of the halo and old disk, respectively, so it is of interest to compare their WD luminosity functions to those from our study.  The WD luminosity functions for a representative globular cluster, M4, and an old open cluster, NGC 6791, are shown in Figure \ref{fig:gc_lf}. From the main-sequence turnoff of the clusters, M4 has been found to have an age of 12.0 $\pm$ 1.4 Gyr \citep{2004ApJS..155..551H}, and NGC 6791 an age of $\sim$ 8~Gyr \citep{2008ApJ...679L..29B}. These ages are comparable to estimates for the age of the inner halo \cite[$12.5^{+1.4}_{-3.4}$ Gyr;][]{2017arXiv170206984K} and that of the thin disk \citep[7.4-8.2 Gyr;][]{2017arXiv170206984K}. The NGVS WD disk luminosity function from Figure \ref{fig:previous_data} is shown in red. Note that the SDSS magnitudes for the disk candidates were converted to the \textit{Hubble Space Telescope} filter system using transformations from \cite{2005PASP..117.1049S} and Lupton (2005)\footnote{\tt\url{ https://www.sdss3.org/dr10/algorithms/sdssUBVRITransform.php}}, and a representative error bar is plotted at the peak of the luminosity function. The luminosity functions are normalized such that the peaks equal 1.

\cite{2008ApJ...678.1279B} note the double peak in the luminosity function of  NGC 6791, which they attribute to unresolved double degenerate binaries. These authors derive a cluster age of $\sim$6 Gyr --- somewhat younger than the age of 8-9 Gyr found from the main-sequence turnoff. \cite{2010Natur.465..194G} argue that this discrepancy can be resolved by considering $^{22}$Ne separation in the cores of the cool WDs, which slows cooling and increases the derived age. Overall, Figure \ref{fig:gc_lf} shows that the NGVS disk luminosity function bears a close resemblance to that of NGC 6791, including the faint peak. The luminosity function for M4 continues to rise beyond the disk turnoff, showing that many of the halo WDs remain undetected, presumably because the faintest objects in the NGVS lack proper motion measurements due to the lack of deep first epoch positions.

\section{Conclusions}
\label{sec:conclusions}

We have used the deep imaging from the NGVS to identify and study WDs within the $\sim$100 deg$^2$ NGVS footprint. WD candidates were identified using three different techniques: (1) a sample of 1209 candidates were selected from the ($g-i$, $u-g$) color-color diagram for point sources based entirely on the NGVS photometry; (2) a sample of 856 candidates were selected from the (NUV$-g$, $g-i$) color-color diagram based on the NGVS and GUViCS surveys; and (3) a sample of 342 candidates were selected from the reduced proper motion diagram derived from NGVS and SDSS. Effective temperatures were calculated by SED fitting of the $u^*giz$ NGVS magnitudes while photometric distances for all candidates were estimated using theoretical color-magnitude relations from \cite{2006AJ....132.1221H}. 

Scale heights computed from these photometric distances were used to separate the WD candidates into possible disk and halo subsamples. Doing so requires accurate photometry since, at the highest WD temperatures probed by the NGVS, small errors in apparent magnitude translate into large errors in distance (i.e., the color-absolute magnitude relation becomes very steep in this temperature regime). 

For the sample of WD candidates selected from the NGVS-SDSS proper motions, a selection based on tangential velocity was used to separate the disk and halo subsamples. Selecting candidates with a tangential velocity larger than 200 km~s$^{-1}$, we find six possible halo WDs. These candidates have relatively high temperatures, resulting in cooling ages more consistent with a disk population. However, the cooling age is highly sensitive to mass, and if their mass is lower than assumed, their ages would increase. This highlights the need for accurate masses when attempting to separate WDs into disk and halo populations. 

The disk and halo subsamples were used to construct luminosity functions that were compared to previous results from \cite{2006AJ....131..571H} and \cite{2011MNRAS.417...93R}. These comparisons are in good agreement. We show that the somewhat steeper luminosity function obtained using the NGVS sample is likely a consequence of the higher mean Galactic latitude ($b \sim 75^\circ$) of the NGVS compared to the SDSS or SuperCOSMOS surveys.

Integrating our WD luminosity function yields a number density for disk WDs of $\phi_d =$ (2.81 $\pm$0.52) $\times 10^{-3}$~pc$^{-3}$, consistent with several previous estimates. The halo number density is  $\phi_h = $ (7.85 $\pm$ 4.55) $\times 10^{-6}$~pc$^{-3}$, or $\sim$0.3\texttt{}\% that of the disk. 

We compared the number of WDs selected by each method to the predictions of two popular Galactic structure models --- Besan\c{c}on \citep{2003A&A...409..523R} and TRILEGAL \citep{2005A&A...436..895G}. This exercise reveals good agreement between the Besan\c{c}on model and the observations after correcting for possible contaminants in the observed sample. The TRILEGAL model, on the other hand, appears to predict too many hot, young WDs compared to the NGVS observations. This discrepancy likely indicates that the IFMR adopted by the TRILEGAL models leads to overly massive WDs \citep{2011MNRAS.411.2770B}. 

We also compare the estimated number of halo WDs in our samples to the predictions of the models. This exercise reveals a discrepancy not only between the models and observations, but also between the models themselves. The discrepancy highlights the lack of firm observational constraints on the input parameters of the models, most notably the IFMR and the observed mass distribution of halo WDs. 

A comparison between the WD luminosity function measured in the NGVS and the observed luminosity function in the old open cluster NGC 6791 shows good agreement. A further comparison between our observed luminosity function and that of the globular cluster M4 indicates that many of the faintest and coolest field halo WDs remain undetected even in the deep NGVS imaging, primarily because proper motions are not yet available to the full depth of the NGVS.

In the future, wide-field photometric and astrometric surveys such as \textit{Gaia} and the Large Synoptic Survey Telescope should dramatically increase our census of WDs, lead to improved measurements of their fundamental parameters, such as mass, and enable more powerful statistical studies of the population of local WDs. For example, \cite{2005MNRAS.360.1381T} used a Monte Carlo simulation to estimate that \textit{Gaia} will be able to identify $>$200,000 disk WDs and $>$1,000 halo WDs to V $\sim$ 21, improving the number of known disk and halo WDs by an order of magnitude. Furthermore, the distance estimates from \textit{Gaia} will yield more accurate masses, allowing for a large statistical study of the mass distributions for WDs belonging to each Galactic component. This study will allow for a more accurate determination of the IFMR.

The ability to study ultracool WDs associated with the halo or old disk will depend on our ability to select candidates at blue wavelengths since the WD cooling sequence turns blueward at low luminosities due to collision-induced opacity from molecular hydrogen in their atmospheres \citep{2003ARA&A..41..465H}. Deep, wide-field $u$-band programs, such as the CFIS/\textit{Luau} survey (Ibata, R. et~al. 2017, in preparation), should provide rich datasets for characterizing the Milky Way's oldest and coolest WDs by selecting candidates for spectroscopic follow-up. Furthermore, \textit{Luau} will provide second epoch positions for stars in the NGVS field, which will result in proper motions for all objects in the NGVS footprint. 
\acknowledgements

We would like to thank the referee for their insightful comments that have improved this paper. This work was supported in part by the Canadian Advanced Network for Astronomical Research (CANFAR), which is made possible by funding from CANARIE under the Network-Enabled Platforms program. This analysis has also used the facilities of the Canadian Astronomy Data Centre (CADC), which is operated by the National Research Council of Canada with the support of the Canadian Space Agency, and data products from the Sloan Digital Sky Survey (SDSS). 
This work was also supported in part by an NSERC Discovery Grant held by D.A.H. E.S. gratefully acknowledges funding by the Emmy Noether program from the Deutsche Forschungsgemeinschaft (DFG). L.B. acknowledges support from NASA grant NNX16AF40G
Funding for the SDSS and SDSS-II has been provided by the Alfred P. Sloan Foundation, the Participating Institutions, the National Science Foundation, the U.S. Department of Energy, the National Aeronautics and Space Administration, the Japanese Monbukagakusho, the Max Planck Society, and the Higher Education Funding Council for England. The SDSS Web Site is http://www.sdss.org/. The SDSS is managed by the Astrophysical Research Consortium for the Participating Institutions. The Participating Institutions are the American Museum of Natural History, Astrophysical Institute Potsdam, University of Basel, University of Cambridge, Case Western Reserve University, University of Chicago, Drexel University, Fermilab, the Institute for Advanced Study, the Japan Participation Group, Johns Hopkins University, the Joint Institute for Nuclear Astrophysics, the Kavli Institute for Particle Astrophysics and Cosmology, the Korean Scientist Group, the Chinese Academy of Sciences (LAMOST), Los Alamos National Laboratory, the Max-Planck-Institute for Astronomy (MPIA), the Max-Planck-Institute for Astrophysics (MPA), New Mexico State University, Ohio State University, University of Pittsburgh, University of Portsmouth, Princeton University, the United States Naval Observatory, and the University of Washington.

\smallskip
{\it Facilities}: CFHT, GALEX, SDSS

\end{document}